\pdfoutput=1
\documentclass[final]{IEEEtran}  
\usepackage{amsthm,amsmath}
\usepackage{epsfig, verbatim, color, amsfonts} 
\usepackage{graphicx}
\title{Communication over Individual Channels}
\author{Yuval Lomnitz, Meir Feder \\
Tel Aviv University, Dept. of EE-Systems \\
Email: \{yuvall,meir\}@eng.tau.ac.il}

\theoremstyle{plain}
\newtheorem{theorem}{Theorem}
\newtheorem{lemma}{Lemma}
\newtheorem{corollary_in_lemma}{Corollary}[lemma]

\theoremstyle{definition}
\newtheorem{definition}{Definition}

\def\vr{\mathbf}
\def\mt{\mathbf}

\def\Pr{\mathrm{Pr}}
\def\Normal{\mathcal{N}}
\def\const{\mathrm{const}}
\def\sign{\mathrm{sign}}
\def\Remp{R_{\mathrm{emp}}}
\def\Ind{\mathrm{Ind}}
\def\half{\frac{1}{2}}

\def\smallminushalf{\left( - \scriptstyle{\half} \right)}
\def\myendofproof{\hspace{\stretch{1}}$\Box$}
\newcommand{\arrowexpl}[1] {\raisebox{-1.0ex}{$\stackrel{\textstyle \longrightarrow}{\scriptscriptstyle #1}$}}
\newcommand{\argmax}[1] {\raisebox{-1.2ex}{$\stackrel{\textstyle \mathrm{argmax}}{\scriptscriptstyle #1}$}}
\newcommand{\argmin}[1] {\raisebox{-1.2ex}{$\stackrel{\textstyle \mathrm{argmin}}{\scriptscriptstyle #1}$}}
\def\RLBONE{R_{\mathrm{LB1}}}
\def\RLBTWO{R_{\mathrm{LB2}}}

\begin{document}
\maketitle

\begin{abstract}
We consider the problem of communicating over a channel for which no mathematical model is specified. We present achievable rates as a function of the channel input and output known a-posteriori for discrete and continuous channels, as well as a rate-adaptive scheme employing feedback which achieves these rates asymptotically without prior knowledge of the channel behavior.
\end{abstract}


\IEEEpeerreviewmaketitle

\section{Introduction}
The problem of communicating over a channel with an individual, predetermined noise sequence which is not known to the sender and receiver was addressed by Shayevitz and Feder \cite{Ofer_BSC} \cite{Ofer_EMP} and Eswaran et al \cite{Eswaran}\cite{Eswaran_conf}. The simple example discussed in \cite{Ofer_BSC} is of a binary channel $y_n=x_n \oplus e_n$ where the error sequence $e_n$ can be any unknown sequence. Using perfect feedback and common randomness, communication is shown to be possible in a rate approaching the capacity of the binary symmetric channel (BSC) where the error probability equals the empirical error probability of the sequence (the relative number of '1'-s in $e_n$). Subsequently both authors extended this model to general discrete channels and modulu-additive channels (\cite{Eswaran}, \cite{Ofer_EMP} resp.) with an individual state sequence, and showed that the empirical mutual information can be attained.

Now we take this model one step further. We consider a channel where no specific probabilistic or mathematical relation between the input and the output is assumed. In order to define positive communication rates without assumptions on the channel, we characterize the achievable rate using the specific input and output sequences, and we term this channel an \textit{individual channel}. This way of treating with unknown channels is different from other concepts of dealing with the problem, such as compound channels and arbitrarily varying channels, in the fact that the later require a specification of the channel model up to some unknown parameters, whereas the current approach makes no a-priori assumptions about the channel behavior. We usually assume the existence of a feedback link in which the channel output or other information from the decoder can be sent back to the encoder. Without this feedback it would not be possible to match the rate of transmission to the quality of the channel so outage would be inevitable.

Although one may not be fully convenient with the mathematical formulation of the problem, there is no question about the reality of this model: this is the only channel model that we know for sure exists in nature. This point of view is similar to the approach used in universal source coding of individual sequences where the goal is to asymptotically attain for each sequence the same coding rate achieved by the best encoder from a model class, tuned to the sequence.

Just to inspire thought, let's ask the following question: suppose the sequence $\{x_i\}_{i=1}^n$ with power
$P = \frac{1}{n}\sum_{i=1}^{n}{x_i^2}$ encodes a message and is transmitted over a continuous real-valued input channel. The output sequence is $\{y_i\}_{i=1}^n$. One can think of $v_i = y_i - x_i$ as a noise sequence and measure its power $N=\frac{1}{n}\sum_{i=1}^{n}{v_i^2}$. Is the rate $R=\half \log \left( 1 + \frac{P}{N} \right)$ which is the Gaussian channel capacity, achievable in this case, under appropriate definitions ?

The way it was posed, the answer to this question would be "no", since this model predicts a rate of $\half$ bit/use for the channel whose output is $\forall i: y_i=0$ which cannot convey any information. However with the slight restatement done in the next section the answer would be "yes".

We consider two classes of individual channels: discrete input and output channels and continuous real valued input and output channels, and two communication models: with feedback and without feedback. In both cases we assume common randomness exists. The case of feedback is of higher interest, since the encoder can adapt the transmission rate and avoid outage. The case of no-feedback is used as an intermediate step, but the results are interesting since they can be used for analysis of semi probabilistic models. The main result is that with a small amount of feedback, a communication at a rate close to the empirical mutual information (or its Gaussian equivalent for continuous channels) can be achieved, without any prior knowledge, or assumptions, about the channel structure.

The paper is organized as follows: in section \ref{sec:overview} we give a high level overview of the results. In section \ref{sec:definitions} we define the model and notation. Section \ref{sec:nonadaptive} deals with communication without feedback where the results pertaining to discrete and continuous case are formalized and proven, and the choice of the rate function and the Gaussian prior for the continuous case is justified. Section \ref{sec:rate_adaptive} deals with the case where feedback is present. After reviewing similar results we state the main result and the adaptive rate scheme that achieves it, and delay the proof to section \ref{sec:analysis}. Here, the error probability and the achieved rate are analyzed and bounded. Section \ref{sec:examples} gives several examples, and section \ref{sec:comments} is dedicated to comments and highlights areas for further study.

\section{Overview of main results}\label{sec:overview}
We start with a high level overview of the definitions and results. The definitions below are conceptual rather than accurate, and detailed definitions follow in the next sections.

A rate function is a function $\Remp : \mathcal{X}^n \times \mathcal{Y}^n \to \mathbb{R}$ of the input and output sequences. In communication without feedback we say a given rate function is achievable if for large block size $n \rightarrow \infty$, it is possible to communicate at rate $R$ and an arbitrarily small error probability is obtained whenever $\Remp$ exceeds the rate of transmission, i.e. whenever $\Remp(\vr{x}, \vr{y}) > R$. In communication with feedback we say a given rate function is achieved by a communication scheme if for large block size $n$, data at rate close to or exceeding $\Remp(\vr{x}, \vr{y})$ is decoded successfully with arbitrarily large probability for every output sequence and almost every input sequence. Roughly speaking, this means that in any instance of the system operation, where a specific $\vr x$ was the input and a specific $\vr y$ was the output, the communication rate had been at least $\Remp(\vr x, \vr y)$. Note that the only statistical assumptions are related to the common randomness, and we consider the rate and error probability \textit{conditioned} on a specific input and output, where the error probability is averaged over common randomness. We say that a rate function $\Remp$ is \textit{an} optimal (but not \textit{the} optimal) function if any $\Remp' \geq \Remp $ which is strictly larger than $\Remp$ at at least one point, is not achievable.

The definition of achievability is not complete without stating the input distribution, since it affects the empirical rate. For example, by setting $\vr x = 0$ one can attain every rate function where $\Remp(0,\vr y)=0$ in a void way, since other $\vr x$ sequences will never appear. Different from classical results in information theory, we do not use the input distribution only as a means to show the existence of good codes: taking advantage of the common randomness we require the encoder to emit input symbols that are random and distributed according to a defined prior
(currently we assume i.i.d. distribution).

The choice of the rate functions is arbitrary in a way: for any pair of encoder and decoder, we can tailor a function $\Remp(\vr{x}, \vr{y})$ as a function equaling the transmitted rate whenever the error probability given the two sequences (averaged over messages and the common randomness) is sufficiently small, and 0 otherwise. However it is clear that there are certain rates which cannot be exceeded uniformly. Our interest will focus on simple functions of the input and output, and specifically in this paper we focus on functions of the instantaneous (zero order) empirical statistics. Extension to higher order models seems technical.

For the discrete channel we show that a rate
\begin{equation}\label{R_emp_discrete}
\Remp = \hat{I} (\vr{x}; \vr{y})
\end{equation}
is achievable with any input distribution $P_X$ where $\hat{I}(\cdot;\cdot)$ denotes the empirical mutual information \cite{Goppa} (see definition in section \ref{sec:definitions}, and Theorems \ref{theorem:discrete_nonadaptive}, \ref{theorem:discrete_adaptive}).
For the continuous (real valued) channel we show that a rate

\begin{equation}\label{R_emp_continuous}
\Remp = \half \log \left( \frac{1}{1-\hat\rho(\vr x, \vr y)^2} \right)
\end{equation}
is achievable with Gaussian input distribution $\Normal(0,P)$, where $\hat \rho$ is the empirical correlation factor between the input and output sequences (see Theorems \ref{theorem:continuous_nonadaptive}, \ref{theorem:continuous_adaptive}).
These results pertain both to the case of feedback and of no-feedback according to the definitions above.

Throughout the current paper we define correlation factor in a slightly non standard way as $\rho = \frac{E(XY)}{\sqrt{E(X^2)E(Y^2)}}$ (that is, without subtracting the mean). This is done only to simplify definitions and derivations, and similar claims can be made using the  correlation factor defined in the standard way. Although the result regarding the continuous case is less tight, we show that this is the best rate function that can be defined by second order moments, and is tight for the Gaussian additive channel (for this channel $\rho^2 = \frac{P}{P+N}$ therefore $\Remp = \half \log \left( 1 + \frac{P}{N}\right)$)

We may now rephrase our example question from the introduction so that it will have an affirmative answer: given the input and output sequences, describe the output by the virtual additive channel with a gain $y_i = \alpha x_i + v_i$, so the effective noise sequence is $v_i = y_i - \alpha x_i$. Chose $\alpha$ so that $\vr v \perp \vr x$, i.e. $\frac{1}{n} \sum_i{v_i x_i}=0$. An equivalent condition is that $\alpha$ minimizes $\lVert \vr v \rVert^2$. The resulting $\alpha$ is the LMMSE coefficient in estimation of $\vr y$ from $\vr x$ (assuming zero mean), i.e. $\alpha = \frac{\vr x ^T \vr y}{\lVert \vr x \rVert^2}$. Define the effective noise power as $N=\frac{1}{n}\sum_{i=1}^{n}{v_i^2}$, and the effective $\textit{SNR} \equiv \frac{\alpha^2 P}{N}$. It is easy to check that $\textit{SNR}=\frac{\hat\rho^2}{1-\hat\rho^2}$ where $\hat\rho=\frac{\vr x ^T \vr y}{\lVert \vr x \rVert \cdot \lVert \vr y \rVert}$ is the empirical correlation factor between $\vr{x}$ and $\vr{y}$. Then according to Eq.(\ref{R_emp_continuous}) the rate $R=\half \log \left( 1 + \textit{SNR} \right)$ is achievable, in the sense defined above. Reexamining the counter example we gave above, in this model if we set $\vr y = 0$ we obtain $\hat\rho=0$ and therefore $\Remp=0$, or equivalently the effective channel has $\vr v = 0$ and $\alpha=0$, therefore $\textit{SNR}=0$ (instead of $\vr v = -\vr x$, $\alpha=1$ and $\textit{SNR}=1$).

As will be seen, we achieve these rates by random coding and universal decoders. For the case of feedback we use iterated instances of rateless coding (i.e. we encode a fixed number of bits and the decision time depends on the channel).
The scheme is able to operate asymptotically with "zero rate" feedback (meaning any positive capacity of the
feedback channel suffices). A similar although more complicated scheme was used in \cite{Eswaran} (see a comparison in the appendix).

Before the detailed presentation we would like to examine the differences between the model used here and two proximate models: the arbitrarily varying channel (AVC) and the channel with individual noise sequence.

In the AVC (see for example \cite{Lapidoth_AVC}\cite{Csiszar_AVC}), the channel is defined by a probabilistic model which includes an unknown state sequence. Constraints on the sequence (such as power, number of errors) may be defined, and the target is to communicate equally well over all possible occurrences of the state sequence. In AVC, the capacity depends on the existence of common randomness and on whether the average or maximum error probability (over the messages) is required to approach $0$, yet when sufficient common randomness is used, the capacities for maximum and average error probability are equal. The notes in \cite{Lapidoth_AVC} regarding common randomness and randomized encoders (see p.2151) are also relevant to our case.

A treatment of AVC-s which is similar in spirit to our results exists in watermarking problems. For example a rather general case of AVC is discussed in \cite{Agarwal_RD}. They consider communication over a black box (representing the attacker) which is only limited to a given level $D$ of distortion according to a predefined metric, but has otherwise a block-wise undefined behavior. They show that it is possible to achieve a rate equal to the rate-distortion function of the input $R_X(D)$, if the black box guarantees a given level of average distortion in high probability. This result is similar to our Theorem \ref{theorem:discrete_nonadaptive}. The remarkable distinction from other results for AVC is that the rate is determined using a constraint on the channel inputs and outputs, rather than the channel state sequence. We note that for the Gaussian additive channel the above result is suboptimal since the rate is $R_X(N)=\half \log(P/N)$ and our results improve this result by using the correlation factor yields rather than the mean squared error. See further discussion of these results in the proof Lemma \ref{lemma:pairwise_discrete} and the discussion following Theorem \ref{theorem:discrete_adaptive}.

Channels with individual noise (or state) sequence are treated by Shayevitz and Feder \cite{Ofer_BSC}\cite{Ofer_EMP} and Eswaran et al \cite{Eswaran}. The probabilistic setting is the same as in the AVC, and the difference is that instead of achieving a uniform (hence worst-case) rate, the target is to achieve a variable rate which depends on the particular sequence of noise, using a feedback link. In this setup, prior constraints on the state sequence can be relaxed. As opposed to AVC where the capacity is well defined, the target rate for each state sequence is determined in a somewhat arbitrary way (since many different constraints on the sequence can be defined). As an example, in the binary channel of \cite{Ofer_BSC}, a rate of 0 would be obtained for the sequence $\vr{e}='01010101...'$ since the empirical error probability is $\half$, although obviously a scheme which favors this specific sequence and achieves a rate of 1 can be designed. On the other hand, with the AVC approach communication over this channel would not be possible without prior constraints on the noise sequence. Channels with individual noise sequence can be thought of as compound-AVCs (i.e. an AVC with unknown parameter, in this case, the constraint). As in AVC, existence of common randomness as well as the definition of error probability affect the achievable rates.

In the individual channel model we use here, since no equation with state sequence connecting the input and output is given, the achievable rates cannot be defined without relating to the channel input. Therefore the definitions of achieved rates depend in a somewhat circular way on the channel input which is determined by the scheme itself. Currently we circumvent this difficulty by constraining the input distribution, as mentioned above.

In many aspects the model used in this paper is more stringent than the AVC and the individual noise sequence models, since it makes less assumptions on the channel, and the error probability is required to be met for (almost) every input and output sequence (rather than on average). In other aspects it is lenient since we may attribute 'bad' channel behavior to the rate rather than suffer an error, therefore the error exponents are better than in probabilistic models. This is further explained in section \ref{sec:discrete_nonadaptive}.

The model we propose suggests a new approach for the design of communication systems. The classical point of view first assumes a channel model and then devises a communication system optimized for it. Here we take the inverse direction: we devise a communication system without assumptions on the channel which guarantees rates depending on channel behavior. This change of viewpoint does not make probabilistic or semi probabilistic channel models redundant but merely suggests an alternative. By using a channel model we can formalize questions relating to optimality such as capacity (single user, networks) and error exponent as well as guarantee a communication rate a-priori. Another aspect is that we pay a price for universality. Even if one considers an individual channel scheme that guarantees asymptotically optimum rates over a large class of channels, it can never consider all possible channels (block-wise), and for a finite block size it will have a larger overhead (a reduction in the amount of information communicated with same error probability) compared to a scheme optimized for the specific channel.

Following our results, the individual channel approach becomes a very natural starting point for determining achievable rates for various probabilistic and arbitrary models (AVC-s, individual noise sequences, probabilistic models, compound channels) under the realm of randomized encoders, since the achievable rates for these models follow easily from the achievable rates for specific sequences, and the law of large numbers. We will give some examples later on.

\section{Definitions and notation}\label{sec:definitions}
\subsection{Notation}\label{sec:notation}
In general we use uppercase letters to denote random variables, respective lowercase letters to denote their sample values and boldface letters to denote vectors, which are by default of length n. However we deviate from this practice when the change of case leads to confusion, and vectors are always denoted by lowercase letters even when they are random variables.

$\lVert \vr x \rVert \equiv \sqrt{\vr x^T \vr x}$ denotes $L_2$ norm. We denote by $P \circ Q$ the product of conditional probability functions e.g. $(P \circ Q)(x,y) = P(x) \cdot Q(y|x)$. A hat ($\hat{\square}$) denotes an estimated value.

We denote the empirical distribution as $\hat{P}$ (e.g. $\hat{P}_{(\vr x, \vr y)}(x,y) \equiv \frac{1}{n} \sum_{i=1}^n {\delta_{(\vr x_i - x), (\vr y_i - y)}}$). The source vectors $\vr x, \vr y$ and/or the variables $x,y$ are sometimes omitted when they are clear from the context. We denote by $\hat{H}(\cdot)$, $\hat{I}(\cdot;\cdot)$, $\hat{\rho}(\cdot;\cdot)$ the empirical entropy, the empirical mutual information
and the empirical correlation factor, which are the respective values calculated for the empirical distribution. All expressions such as $\hat H(\vr x)$, $\hat H(\vr x | \vr y)$, $\hat I (\vr x; \vr y)$, $\hat I (\vr x; \vr y | \vr z)$, $\hat I (\vr x; \vr y | \vr z = z_0)$ are interpreted as their respective probabilistic counterparts $H(X)$, $H(X|Y)$, $I (X;Y)$, $I (X;Y|Z)$, $I (X;Y|Z=z_0)$ where $(X,Y,Z)$ are random variables distributed according to the empirical distribution of the vectors $\hat{P}_{(\vr x, \vr y, \vr z)}$, or equivalently are defined as a random selection of an element of the vectors i.e. $(X,Y,Z)=(x_i, y_i, z_i), i \sim \mathrm{U}\{1,\ldots,n\}$. It is clear from this equivalence that relations on entropy and mutual information (e.g. positivity, chain rules) are directly translated to relations on their empirical counterparts.

We apply superscript and subscript indices to vectors to define subsequences in the standard way, i.e. $\vr x_i^j \equiv (x_i, x_{i+1}, ... , x_j)$, $\vr x^i \equiv \vr x_1^i$

We denote $I(P,W)$ the mutual information $I(X;Y)$ when $(X,Y) \sim P(x)\cdot W(y|x)$. $\mathrm{U}(A)$ denotes a uniform distribution over the set $A$. $Ber(p)$ denotes the Bernoulli distribution, and $h_b(p) \equiv H(Ber(p)) = -p \log p - (1-p) \log (1-p) $ denotes the binary entropy function. The indicator function $\Ind(E)$ where $E$ is a set or a probabilistic event is defined as $1$ over the set (or when the event occurs) and $0$ otherwise.

The functions $\log(\cdot)$ and $\exp(\cdot)$ as well as information theoretic quantities $H(\cdot), I(\cdot;\cdot), D(\cdot || \cdot)$ refer to the same, unspecified base. We use the term "information unit" as the unit of these quantities (equals $\frac{1}{\log(2)}$ bits).

The notation $f_n = O(g_n)$ and $f_n < O(g_n)$ (or equivalently $O(f_n) = O(g_n)$ and $O(f_n) < O(g_n)$) means $\frac{f_n}{g_n} \arrowexpl{n \to \infty} \const > 0$ and $\frac{f_n}{g_n} \arrowexpl{n \to \infty} 0$ respectively.

Throughout this paper we use the term "continuous" to refer to the continuous \emph{real valued} channel $\mathbb{R} \to \mathbb{R}$, although this definition does not cover all continuous input - continuous output channels. By the term "discrete" in this paper we always refer to finite alphabets (as opposed to countable ones).

\subsection{Definitions}\label{sec:definitions}
\begin{definition}[Channel]
A channel is defined by a pair of input and output alphabets $\mathcal{X,Y}$, and denoted $\mathcal{X} \to \mathcal{Y}$
\end{definition}

\begin{definition}[Fixed rate encoder, decoder, error probability]
A randomized block encoder and decoder pair for the channel $\mathcal{X \to Y}$ with block length $n$ and rate $R$ without feedback is defined by a random variable $S$ distributed over the set $\mathcal{S}$, a mapping $\phi : \{1,2,\ldots\exp(nR)\} \times \mathcal{S} \to \mathcal{X}^n$ and a mapping $\bar\phi : \mathcal{Y}^n \times \mathcal{S} \to \{1,2,
\ldots\exp(nR)\}$. The error probability for message $w \in \{1,2,\ldots\exp(nR)\}$ is defined as
\begin{equation}
P_e^{(w)}(\vr x, \vr y) = \Pr \left(\bar\phi(\vr y, S) \neq w \big\vert \phi(w,S) = \vr x \right)
\end{equation}

where for $\vr x$ such that the condition cannot hold, we define $P_e^{(w)}(\vr x, \vr y)=0$.
\end{definition}

Note that the encoder rate must pertain to a discrete number of messages $\exp(nR) \in \mathbb{Z}_+$, but the empirical rates defined in the following theorems may be any positive real numbers.

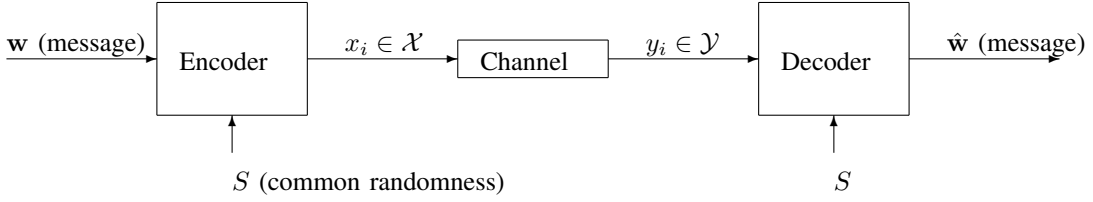
\begin{figure*}[t]
\setlength{\unitlength}{1mm}
\hspace{\stretch{1}}
\begin{picture}(140,30)
\put(23,16){Encoder}\put(20,10){\line(1,0){20}}\put(40,10){\line(0,1){15}}\put(40,25){\line(-1,0){20}}\put(20,25){\line(0,-1){15}}
\put(63,16){Channel}\put(60,15){\line(1,0){20}}\put(80,15){\line(0,1){5}}\put(80,20){\line(-1,0){20}}\put(60,20){\line(0,-1){5}}
\put(103,16){Decoder}\put(100,10){\line(1,0){20}}\put(120,10){\line(0,1){15}}\put(120,25){\line(-1,0){20}}\put(100,25){\line(0,-1){15}}
\put(0,17.5){\vector(1,0){20}}\put(0,18.5){$\vr w$ (message)}
\put(40,17.5){\vector(1,0){20}}\put(45,18.5){$x_i \in \mathcal{X}$}
\put(80,17.5){\vector(1,0){20}}\put(85,18.5){$y_i \in \mathcal{Y}$}
\put(120,17.5){\vector(1,0){20}}\put(125,18.5){$\hat{\vr w}$ (message)}
\put(30,5){\vector(0,1){5}}\put(30,0){$S$ (common randomness)}
\put(110,5){\vector(0,1){5}}\put(110,0){$S$}
\end{picture}
\hspace{\stretch{1}}
\caption{Non rate adaptive encoder-decoder pair without feedback}\label{fig:system_non_adaptive}
\end{figure*}

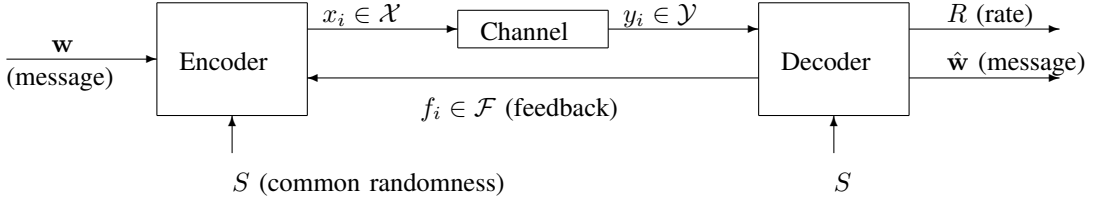
\begin{figure*}[t]
\setlength{\unitlength}{1mm}
\hspace{\stretch{1}}
\begin{picture}(140, 30)
\put(23,16){Encoder}\put(20,10){\line(1,0){20}}\put(40,10){\line(0,1){15}}\put(40,25){\line(-1,0){20}}\put(20,25){\line(0,-1){15}}
\put(63,20){Channel}\put(60,19){\line(1,0){20}}\put(80,19){\line(0,1){5}}\put(80,24){\line(-1,0){20}}\put(60,24){\line(0,-1){5}}
\put(103,16){Decoder}\put(100,10){\line(1,0){20}}\put(120,10){\line(0,1){15}}\put(120,25){\line(-1,0){20}}\put(100,25){\line(0,-1){15}}
\put(0,17.5){\vector(1,0){20}}\put(6,18.5){$\vr w$}\put(0,14){(message)}
\put(40,21.5){\vector(1,0){20}}\put(42,22.5){$x_i \in \mathcal{X}$}
\put(80,21.5){\vector(1,0){20}}\put(82,22.5){$y_i \in \mathcal{Y}$}
\put(100,15){\vector(-1,0){60}}\put(55,10){$f_i \in \mathcal{F}$ (feedback)}
\put(120,21.5){\vector(1,0){20}}\put(125,22.5){$R$ (rate)}
\put(120,15){\vector(1,0){20}}\put(125,16){$\hat{\vr w}$ (message)}
\put(30,5){\vector(0,1){5}}\put(30,0){$S$ (common randomness)}
\put(110,5){\vector(0,1){5}}\put(110,0){$S$}
\end{picture}
\hspace{\stretch{1}}
\caption{Rate adaptive encoder-decoder pair with feedback}\label{fig:system_adaptive}
\end{figure*}

\begin{definition}[Adaptive rate encoder, decoder, error probability]
A randomized block encoder and decoder pair for the channel $\mathcal{X \to Y}$ with block length $n$, adaptive rate and feedback is defined as follows:
\begin{itemize}
\item The message $\vr w$ is expressed by the infinite sequence $\vr w_1^{\infty} \in \{0,1\}^{\infty}$
\item The common randomness is defined as a random variable $S$ distributed over the set $\mathcal{S}$
\item The feedback alphabet is denoted $\mathcal{F}$
\item The encoder is defined by a series of mappings $ x_k = \phi_k(\vr w, s, \vr f^{k-1})$ where $\phi_k : \{0,1\}^{\infty} \times \mathcal{S} \times \mathcal{F}^{k-1} \to \mathcal{X}$.
\item The decoder is defined by the feedback function $\varphi_k: \mathcal{Y}^{k-1} \times \mathcal{S} \to \mathcal{F}$, the decoding function $\bar\phi : \mathcal{Y}^n \times \mathcal{S} \to \{0,1\}^{\infty}$ and the rate function $r : \mathcal{Y}^n \times \mathcal{S} \to \mathbb{R}^+$ (where the rate is measured in bits), applied as follows:

\begin{eqnarray}
f_k &=& \varphi_k(\vr y^k, S) \\
\hat {\vr w} &=& \bar\phi (\vr y, S) \\
R &=& r (\vr y, S)
\end{eqnarray}
\end{itemize}

The error probability for message $\vr w$ is defined as

\begin{equation}
P_e^{(\vr w)}(\vr x, \vr y) = \Pr \left({\hat {\vr w}}_1^{\lceil nR \rceil} \neq  {{\vr w}}_1^{\lceil nR \rceil} \big\vert \vr x, \vr y \right)
\end{equation}

In other words, a recovery of the first $\lceil nR \rceil$ bits by the decoder is considered a successful reception. For $\vr x$ such that the condition cannot hold, we define $P_e^{(\vr w)}(\vr x, \vr y)=0$.  The conditioning on $\vr y$ is mainly for clarification, since it can be treated as a fixed vector. This system is illustrated in figure \ref{fig:system_adaptive}.

\end{definition}

Note that if we are not interested in limiting the feedback rate, and perfect feedback can be assumed, the definition of feedback alphabet and feedback function is redundant (in this case $\mathcal{F} = \mathcal{Y}$ and $f_k=y_k$). The model in which the decoder determines the transmission rate is lenient in the sense that it gives the flexibility to exchange rate for error probability: the decoder may estimate the error probability and decrease it by reducing the decoding rate. In the scheme we discuss here the rate is determined during reception, but it's worth noting in this context the posterior matching scheme \cite{Ofer_Posterior_analysis} for the known memoryless channel. In this scheme the message is represented as a real number $\theta \in [0,1)$ and the rate for a given error probability $P_e$ can be determined \emph{after} the decoding by calculating $\Pr(\theta \vert \vr y)$ and finding the smallest interval with probability at least $1-P_e$.

\section{Communication without feedback}\label{sec:nonadaptive}
In this section we show that the empirical mutual information (in the discrete case) and its Gaussian counterpart (in the continuous case) are achievable in the sense defined in the overview. For the continuous case we justify the choice of the Gaussian distribution as the one yielding the maximum rate function that can be defined by second order moments.

\subsection{The discrete channel without feedback}\label{sec:discrete_nonadaptive}
The following theorem formalizes the achievability of rate $\hat{I} (\vr{x}; \vr{y})$ without feedback:

\begin{theorem}[Non-adaptive, discrete channel]\label{theorem:discrete_nonadaptive}
Given discrete input and output alphabets $\mathcal{X,Y}$, for every $P_e>0$, $\delta>0$, prior $Q(x)$ over $\mathcal{X}$ and rate $R>0$ there exists $n$ large enough and a random encoder-decoder pair of rate $R$ over block size $n$, such that the distribution of the input sequence is $\vr{x} \sim Q^n$ and the probability of error for any message given an input sequence $\vr{x} \in \mathcal{X}^n$ and output sequence $\vr{y} \in \mathcal{Y}^n$ is not greater than $P_e$ if $\hat I (\vr{x},\vr{y})> R + \delta$.
\end{theorem}

Theorem \ref{theorem:discrete_nonadaptive} follows almost immediately from the following lemma, which is proven in the appendix using simple a calculation based on the method of types \cite{MethodOfTypes}:

\begin{lemma}\label{lemma:pairwise_discrete}
For any sequence $\vr y \in \mathcal{Y}^n$ the probability of a sequence $\vr x \in \mathcal{X}^n$ drawn independently according to $Q^n$ to have $\hat{I}(\vr x; \vr y) \geq t$ is upper bounded by:
\begin{equation}
Q^n \left( \hat{I}(\vr x; \vr y) \geq t \right) \leq \exp \left( -n \left(t - \delta_n \right) \right)
\end{equation}
where $\delta_n = |\mathcal{X}||\mathcal{Y}|\frac{\log(n+1)}{n} \to 0$.
\end{lemma}
Following notations in \cite{MethodOfTypes}, $Q^n(A)$ denotes the probability of the event $A$ or equivalently the set of sequences $A$ under the i.i.d. distribution $Q^n$. Remarkably this bound does not depend on $Q$.

To prove Theorem \ref{theorem:discrete_nonadaptive}, the codebook $\{\vr x_m\}_{m=1}^{\exp(nR)}$ is randomly generated by i.i.d. selection of its $L = \exp(nR) \cdot n$ letters, so that the common randomness $S \in \mathcal{X}^{L}$ may be defined as the codebook itself and is distributed $Q^{L}$. The encoder sends the $w$-th codeword, and the decoder uses maximum mutual information decoding (MMI) i.e. chooses:
\begin{equation} \hat{w} = \bar\phi (\vr y,\{\vr x_m\}) = \argmax{m} \left[ \hat I (\vr x_m ; \vr y) \right] \end{equation}
where ties are broken arbitrarily. By Lemma \ref{lemma:pairwise_discrete}, the probability of error is bounded by:
\begin{multline}
P_e^{(w)}(\vr x_w, \vr y) \leq \Pr \left\{ \bigcup_{m \neq w} \left( \hat I (\vr x_m ; \vr y) \geq \hat I (\vr x_w ; \vr y) \right) \right\} \leq \\
\leq \exp(nR) \exp \left( -n \left(\hat I (\vr x_w ; \vr y) - \delta_n \right) \right)
=\\= \exp \left( -n \left(\hat I (\vr x_w ; \vr y) - R  - \delta_n \right) \right)
\end{multline}
For any $\delta$ there is $n$ large enough such that $\frac{-\log(P_e)}{n} + \delta_n < \delta$. For this $n$, whenever
$\hat I (\vr x ; \vr y) > R + \delta$ we have
\begin{equation}
P_e^{(w)}(x,y) \leq \exp \left( -n \left(\delta  - \delta_n \right) \right) < P_e
\end{equation}
which proves the theorem. \myendofproof

Note that the MMI decoder used here is a popular universal decoder (see \cite{Goppa}\cite{MethodOfTypes}\cite{Tchamkerten}), and was shown to achieve the same error exponent as the maximum likelihood decoder for fixed composition codes. The error exponent obtained here is better than the classical error exponent (slope of -1), and the reason is that the behavior of the channel is known, and therefore no errors occur as result of non-typical channel behavior. Comparing for example with the derivation of the random coding error exponent for the probabilistic DMC based on the method of types (see \cite{MethodOfTypes}), in the later the error probability is summed across all potential "behaviors" (conditional types) of the channel accounting for their respective probabilities (resulting in one behavior, usually different from the typical behavior, dominating the bound), while here the behavior of the channel (the conditional distribution) is fixed, and therefore the error exponent is better. This is not necessarily the best error exponent that can be achieved (see \cite{Tchamkerten}\cite{Burnashev} which discuss error exponent with random decision time and feedback for probabilistic and compound models).

Note that the empirical mutual information is always well defined, even when some of the input and output symbols do not appear in the sequence, since at least one input symbol and one output symbol always appear. For the particular case of empirical mutual information measured over a single symbol, the empirical distributions become unit vectors (representing constants) and their mutual information is 0.

In this discussion we have not dealt with the issue of choosing the prior $Q(x)$. Since the channel behavior is unknown it makes sense to choose the maximum entropy, i.e. the uniform, prior which was shown to obtain a bounded loss from capacity \cite{Shulman_Prior}.

\subsection{The continuous channel without feedback}\label{sec:continuous_nonadaptive}
When turning to define empirical rates for the real valued alphabet case, the first obstacle we tackle is the definition of the empirical mutual information. A potential approach is to use discrete approximations. We only briefly describe this approach since it is somewhat arbitrary and less elegant than in the discrete case. The main focus is on empirical rates defined by the correlation factor. Although the later approach is pessimistic and falls short of the mutual information for most channels, it is much simpler and elegant than discrete approximations. We believe this approach can be further extended to obtain results closer to the (probabilistic) mutual information.

\subsubsection{Discrete approximations}\label{sec:continuous_by_quantization}
Define the continuous input and output alphabets $\mathcal{X},\mathcal{Y}$. Suppose $Q$ is an arbitrary (continuous) prior. Define input and output quantizers to discrete alphabets $A_n : \mathcal{X} \to \mathcal{\tilde{X}}_n$ and $B_n : \mathcal{Y} \to \mathcal{\tilde{Y}}_n$ where $\mathcal{\tilde{X}}_n$, $\mathcal{\tilde{Y}}_n$ are discrete alphabets of growing size, chosen to grow slowly enough so that $\delta_n = |\mathcal{\tilde{X}}_n||\mathcal{\tilde{Y}}_n|\frac{\log(n+1)}{n} \arrowexpl{n \to \infty} 0$. Define the empirical mutual information between continuous vectors as the empirical mutual information between their quantized versions (quantized letter by letter):
\begin{equation}\hat I_{A,B}(\vr x, \vr y) \equiv \hat I(A_n(\vr x), B_n(\vr y))\end{equation}
Then based on Lemma \ref{lemma:pairwise_discrete}, by using a random codebook drawn according to $Q$ and applying a maximum mutual information decoder using the above definition, we could asymptotically achieve the rate function $\Remp = \hat I_{A,B}(\vr x, \vr y)$ based on the definitions of Theorem \ref{theorem:discrete_nonadaptive}. The main issue with this approach is that determining $A_n, B_n$ is arbitrary, and especially $B_n$ is difficult to define when the output range is unknown. Therefore in the following we focus on the suboptimal approach using the correlation factor.

\subsubsection{Choosing the input distribution and rate function}\label{sec:choosing_continunous_rate_function}
First we justify our choice of the Gaussian input distribution and the aforementioned rate function. We take the point of view of a compound (probabilistic, unknown) channel. If a rate function cannot be attained for compound channel model, it cannot be attained also in the more stringent individual model. It is well known that for a memoryless  additive noise channel with constraints on the transmit power and noise variance, the Gaussian noise is the worst noise when the prior is Gaussian, and the Gaussian prior is the best prior when the noise is Gaussian. Thus by choosing a Gaussian prior we choose the best prior for the worst noise, and can we guarantee the mutual information will equal, at least, the Gaussian channel capacity. See the "mutual information game" (problem 9.21) in \cite{Cover}. For the additive noise channel \cite{ErezGaussian} shows the loss from capacity when using Gaussian distribution is limited to $\half$ a bit. However the above is true only for additive noise channels. For the more general where no additivity is assumed case we show below (Lemma \ref{lemma:R_rho_best}) that the rate function $R=-\half\log(1-\rho^2)$ is the best rate function that can be defined by second order moments, and attained universally. Of course, this proof merely supplies the motivation to use a Gaussian distribution and does not rid us from the need to prove this rate is achievable for specific, individual sequences.

\begin{lemma}\label{lemma:gaussian_mi_bound}
Let X,Y be two continuous random variables with correlation factor $\rho \equiv \frac{E(XY)}{\sqrt{E(X^2)E(Y^2)}}$, where $X$ is Gaussian $X \sim \Normal(0,P)$. Then
$I(X;Y)\geq -\half \log(1 -\rho^2)$
\end{lemma}

\begin{corollary_in_lemma}\label{corollary1_gaussian_mi_bound}
Equality holds iff X,Y are jointly Gaussian \end{corollary_in_lemma}

\begin{corollary_in_lemma}\label{corollary2_gaussian_mi_bound}
The lemma does not hold for general X (not Gaussian) \end{corollary_in_lemma}


The proof is given in the appendix. Note that $-\half \log(1 -\rho^2)$ is the mutual information of two Gaussian r.v-s (\cite{Cover}, example 8.5.1). Also note the relation to Theorem 1 in \cite{Hassibi} dealing with an additive channel with uncorrelated, but not necessarily independent noise. The following lemma justifies our selection of $R(\rho) = -\half \log(1 - \rho^2)$:

\begin{lemma}\label{lemma:R_rho_best}
Let $Q(x)$ be an input prior, $W(y|x)$ be an unknown channel, $\Lambda(Q,W)$ be the correlation matrix $\Lambda \equiv E \left( \substack{X \\ \\ Y} \right) \left( \substack{X \\ \\ Y} \right)^T$ between $X,Y$ induced by the joint probability $Q \circ W$ and $\rho(Q,W)$ be the correlation factor induced by $Q,W$ ($\rho = \frac{\Lambda_{12}}{ \sqrt{\Lambda_{11}\Lambda_{22}}}$). We say a function $R(\Lambda)$ is an attainable second order rate function if there exists a $Q(x)$ such that for every channel $W(y|x)$ inducing correlation $\Lambda$ the mutual information is at least $R(\Lambda)$ (in other words can carry the rate $R(\Lambda)$). Then $R(\Lambda) = -\half \log(1 - \rho^2)$ is the largest attainable second order rate function.

Alternatively this can be stated as:
\begin{equation} R(\Lambda) \equiv \max_{Q} \min_{W: \Lambda(Q,W) = \Lambda} I(Q,W) = -\half \log(1 - \rho^2) \end{equation}

\end{lemma}

\textit{Proof of lemma \ref{lemma:R_rho_best}}:
$R(\Lambda) = -\half \log(1 - \rho^2)$ is attainable by selecting an input prior $Q = \Normal(0,\sigma_x^2)$ and by lemma \ref{lemma:gaussian_mi_bound} the mutual information is at least $R(\Lambda)$ for all channels. $R(\Lambda)$ is the maximum attainable function since by writing the condition of the lemma for the additive white gaussian noise (AWGN) channel $W^*$ (a specific choice of $W$) and any $Q$, we have $R(\Lambda) \leq I(Q,W^*) \leq  I(\Normal(0,E_P(X^2),W^*)) = -\half \log(1-\rho^2)$, where the inequalities follow from the conditions of the lemma on $R$ and from the fact the Gaussian prior achieves the AWGN capacity.

\subsubsection{Communication scheme for the empirical channel (without feedback)}\label{sec:nonadaptive_continuous_achievability}
The following theorem is the analogue of Theorem \ref{theorem:discrete_nonadaptive} where the expression $-\half \log(1 - \rho^2)$ (interpreted as the Gaussian effective mutual information) plays the role of mutual information.

\begin{theorem}[Non-adaptive, continuous channel]\label{theorem:continuous_nonadaptive}
Given the channel $\mathbb{R} \to \mathbb{R}$ for every $P_e>0$, $\delta>0$, power $P>0$ and rate $R>0$ there exists
$n$ large enough and a random encoder-decoder pair of rate $R$ over block size $n$, such that the distribution of the input sequence is $\vr{x} \sim \Normal^n(0,P)$ and the probability of error for any message given an input sequence $\vr{x}$ and output sequence $\vr{y}$ with empirical correlation $\hat\rho$ is not greater than $P_e$ if $\Remp = \half \log \left( \frac{1}{1 - \hat\rho^2} \right)  > R + \delta$
\end{theorem}

As before, the theorem will follow easily from the following lemma, proven in the appendix.

\begin{lemma}\label{lemma:pairwise_continuous}
Let $\vr x, \vr y \in \mathbb{R}^n$ be two sequences, and $\hat\rho \equiv \frac{\vr x^T \vr y}{\lVert \vr x \rVert \lVert \vr y \rVert }$ be the empirical correlation factor. For any $\vr y$, the probability of $\vr x$ drawn according to $\Normal^n(0,\sigma_x^2)$ to have $|\hat\rho| \geq t$ is bounded by:
\begin{equation} \Pr ( |\hat\rho| \geq t ) \leq 2 \exp\left(-(n-1) R_2(t)\right) \end{equation}
where
\begin{equation} R_2(t) \equiv \half \log \left( \frac{1}{1 - t^2} \right) \end{equation}
\end{lemma}

To prove Theorem \ref{theorem:continuous_nonadaptive}, the codebook $\{\vr x_m\}_{m=1}^{\exp(nR)}$ is randomly generated by Gaussian i.i.d. selection of its $L = \exp(nR) \cdot n$ letters, and the common randomness $S \in \mathcal{X}^{L}$ is defined as the codebook itself and is distributed $\Normal^{L}(0,P)$. The encoder sends the $w$-th codeword, and the decoder uses maximum empirical correlation decoder i.e. chooses:
\begin{equation}
\hat{w} = \bar\phi (\vr y,\{\vr x_m\}) = \argmax{m}  \rvert \hat \rho (\vr x_m ; \vr y) \lvert  =
 \argmax{m} \left[ \frac{\lvert \vr x_m^T \vr y \rvert }{\lVert \vr x_m \rVert} \right]
\end{equation}
where ties are broken arbitrarily. By Lemma \ref{lemma:pairwise_continuous}, the probability of error is bounded by:
\begin{multline}
P_e^{(w)}(\vr x_w, \vr y) \leq \Pr \left\{ \bigcup_{m \neq w} \left( |\hat \rho (\vr x_m ; \vr y)| \geq |\hat \rho (\vr x_w ; \vr y)| \right) \right\} \leq \\
\leq \exp(nR) \cdot 2 \exp\left(-(n-1) R_2(\hat \rho (\vr x_w ; \vr y)))\right)
=\\= 2 \exp(R) \cdot \exp \left( -(n-1) \left(R_2(\hat \rho) - R\right) \right)
\end{multline}

Choosing $n$ large enough so that $\frac{1}{n-1} \left( R+\log \left(\frac{2}{P_e}\right) \right) \leq \delta$
 (where $P_e$ is from Theorem \ref{theorem:continuous_nonadaptive}) we have that when
$R_2(\hat \rho) > R +  \delta$:
\begin{equation}
P_e^{(w)}(x,y) \leq  2 \exp(R) \cdot \exp \left( -(n-1) \delta  \right) \leq P_e
\end{equation}
which proves the theorem.
\myendofproof

A note is due regarding the definition of $\hat\rho$ in singular cases where $\vr x$ or $\vr y$ are 0. The limit of $\hat\rho$ as $\vr y \to 0$ is undefined (the directional derivative may take any value in [0,1]), however for consistency we define $\hat\rho=0$ when $\vr y=0$. Since $\vr x$ is generated from a Gaussian distribution we do not worry about the event $\vr x=0$ since the probability of this event is 0.

It's worth spending a few words on the connections between the receivers used for the discrete and the continuous cases. Since the mutual information between two Gaussian r.v-s is $-\half \log(1 - \rho^2)$, one can think of this value as a measure of mutual information under Gaussian assumptions. Thus, using this metric as an effective mutual information, since the mutual information is an increasing function of $|\rho|$ the MMI decoder becomes a maximum empirical correlation decoder. On the other hand, the receiver we used can be identified as the GLRT (generalized maximum likelihood ratio test) for the AWGN channel $Y=\alpha X + \Normal(0,\sigma^2)$ with $\alpha$ an unknown parameter, resulting from maximizing the likelihood of the codeword and the channel simultaneously:
\begin{multline}
\hat w = \argmax{\vr x_m} \max_{\alpha} \log \Pr(\vr y|\vr x;\alpha)
=\\= \argmin{m} \min_{\alpha} \lVert \vr y - \alpha \vr x_m \rVert^2 = \argmax{m} \frac{(\vr x_m^T \vr y)^2}{\lVert \vr x_m \rVert^2}
=\\=  \argmax{m} \left[ \hat\rho^2(\vr x_m, \vr y) \right]
\end{multline}
The choice of the GLRT is motivated by considering the individual channel as an effective additive channel with unknown gain (as presented in section \ref{sec:overview}), combined with the fact Gaussian noise is the worse. For discrete memoryless channels it is easy to show that the GLRT (where the group of channels consists of all DMC-s) is synonymous with the MMI decoder (see \cite{Lapidoth_AVC}). Thus, we can identify the two decoders as GLRT decoders, or equivalently as variants of MMI decoders. In the sequel we sometimes use the term "empirical mutual information" in a broad sense that includes also the metric $-\half \log(1 - \hat\rho^2)$.

Regarding the receiver required to obtain the rates of Theorem \ref{theorem:continuous_nonadaptive}, it is interesting to consider the simpler maximum projection receiver $\argmax{\vr x_m} |\vr x_m^T \vr y|$. This receiver seems to differ from the maximum correlation receiver only in the term $\lVert \vr x_m \rVert$ which is nearly constant for large $n$ due to the law of large numbers. However surprisingly, the maximum rate achievable with the projection receiver is only $\half \hat\rho^2$ as can be shown by a simple calculation equivalent to Lemma \ref{lemma:pairwise_continuous} (simpler, since $z = \vr x^T \vr y$ is Gaussian). The reason is that when $\vr x$ is chosen independently of $\vr y$, a large value of the projection (non typical event) is usually created by a sequence with power significantly exceeding the average (another non typical event). When one non-typical event occurs there is no reason to believe the sequence is typical in other senses thus the approximation $\lVert \vr x_m \rVert \approx \sqrt {n P}$ is invalid. The correlation receiver normalizes by the power of $\vr x$ and compensates this effect. An alternative receiver which yields the rates of Theorem \ref{theorem:continuous_nonadaptive} and is similar to the AEP receiver looks for the codeword with the maximum absolute projection subject to power limited to $\frac{1}{n} \lVert \vr x_m \rVert^2 < P+\epsilon$. This can be shown by Sanov theorem \cite{MethodOfTypes} or by using the Chernoff bound. The maximum correlation receiver was chosen because of its elegance and the simplicity of the proof of Lemma \ref{lemma:pairwise_continuous}.

Combining this lemma with the law of large numbers provides a simple proof for the achievability of the AWGN capacity ($\half\log(1+\textit{SNR})$), which uses much simpler mechanics than the popular proof based on AEP or error exponents. This receiver has the technical advantage, compared to the AEP receiver, that it does not declare an error for codewords which have power deviating from the nominal power. This technical advantage is important in the context of rateless decoding since the power condition needs to be re-validated each symbol, thus increasing its contribution to the overall error probability.

Lapidoth \cite{Lapidoth_Nearest} showed that the nearest neighbor receiver achieves a rate equal to the Gaussian capacity $\half \log (1 + P/N)$ over the additive channel $Y=X+V$ with arbitrary noise distribution (with fixed noise power). This result parallels the result that the random code capacity of the AVC $Y=X+V$ with a power constraint on $V$ equals the Gaussian capacity \cite{Hughes_AVC} (this stems directly from the characterization of the random code capacity of the AVC as $\max_{P_X(x)}\min_{P_S(s)} I(X;Y)$, cf.\cite{MethodOfTypes} Eq.(V.4)). Our result is stronger since it does not assume the channel is additive (nor any fixed behavior), but considering the former results it is not surprising, if one assumes (1) that any channel can be modeled as $Y=\alpha X+V$ with $V \perp X$, (2) that the dependence of $V$ on $X$ does not increase the error probability due to orthogonality (see \cite{Hassibi}) and (3) that the loss from the single unknown parameter $\alpha$ is asymptotically small.

Another related result is Agarwal et al's \cite{Agarwal_RD} result that it is possible to communicate with a rate approaching the rate-distortion function $R_X(D)$ over an arbitrarily varying channel with unknown block-wise behavior satisfying a distortion constraint $\hat E d(\vr x, \vr y) \leq D$ in high probability. This relation is further discussed in the proof of Lemma \ref{lemma:pairwise_discrete}. Their result is similar to ours in the fact they define the rate in terms of the input and output alone. The result is similar to obtaining the rate function $\Remp \approx R_X(\hat E d(\vr x, \vr y))$ in the sense of Theorems \ref{theorem:discrete_nonadaptive},\ref{theorem:continuous_nonadaptive}. However their result is not tight even for the Gaussian channels: for the gaussian channel $Y=X+V$ with noise $V$ limited to power $N$ and the Gaussian prior $X \sim \Normal(0,P)$ this rate function equals $R_X(N) = \half \log \left( \frac{P}{N} \right)$ which is smaller than this channel capacity, whereas with Theorem \ref{theorem:continuous_nonadaptive} we would obtain $\half \log \left( 1 + \frac{P}{N} \right)$. Agarwal's result is tight in the sense that this is the maximum rate that can be guaranteed given this distortion. There exists a channel with the same distortion $N$ whose capacity is only $\half \log \left( \frac{P}{N} \right)$: the channel $Y = \alpha  X + \beta V$ with $\alpha = \beta^2 = 1-\frac{N}{P}$. The reason for the sub-optimality of the result is that the squared distance between the input and output, in contrast with the correlation factor, does not yield a tight representation of all memoryless linear Gaussian channels (in the sense of Lemma \ref{lemma:R_rho_best}).

\section{Communication with feedback}\label{sec:rate_adaptive}
\subsection{Overview and background}\label{sec:rate_adaptive_overview}
In this section we present the rate-adaptive counterparts of Theorems \ref{theorem:discrete_nonadaptive}, \ref{theorem:continuous_nonadaptive}, and the scheme achieving them. The proof is delayed to the next section. The scheme we use in order to adaptively attain these rates is by iterating a rateless coding scheme. In other words, in each iteration we send a fixed number of bits $K$, by transmitting symbols from an $n$ length codebook, until the receiver has enough information to decode. Then, the receiver sends an indication that the block is over and a new block starts.

Before developing the details we give some background regarding the evolution of rateless codes, and the differences between the proposed techniques. The earliest work is of Burnashev \cite{Burnashev} who showed that for known channels, using feedback and a random decision time (i.e. decision time which depends on the channel output) yields an improved error exponent, which is attained by a 3 step protocol (best described in \cite{Tchamkerten}) and shown to be optimal. Shulman  \cite{Shulman} proposed to use random decision time as a means to deal with sending common information over broadcast channels (static broadcasting), and for unknown compound channels (which are treated as broadcast). In this scheme later described as "rateless coding" (or Incremental Redundancy Hybrid ARQ) a codebook of $\exp(K)$ infinite sequences is generated, and the sequence representing the message is sent to the receiver symbol by symbol, until the receiver decides to decode (and turn off, in case of a broadcast channel). Tchamkerten and Telatar \cite{Tchamkerten}  connect the two results by showing that for some, but not all compound channels Burnashev error exponent can be attained universally using rateless coding and the 3 step protocol. Eswaran, Sarwate, Sahai and Gastpar \cite{Eswaran} used iterated rateless coding to achieve the mutual information related to the empirical noise statistics on channels with individual noise sequences. The scheme we use here is most similar to the one used in \cite{Eswaran} but less complicated. We do not use training symbols to learn the channel in order to decide on the decoding time but rely on the mutual information itself as the criterion (based on Lemmas \ref{lemma:pairwise_discrete},\ref{lemma:pairwise_continuous}) and the partitioning into blocks and the decision rules are simpler. The result in \cite{Eswaran} is an extension of a result in \cite{Ofer_BSC} regarding the binary channel to general discrete channels with individual noise sequence. The original result in \cite{Ofer_BSC} was obtained not by rateless codes but by a successive estimation scheme \cite{Ofer_Posterior} which is a generalization of the Horstein \cite{Horstein} and Schalkwijk-Kailath \cite{Schalkwijk} schemes. The same authors extend their results to discrete channels \cite{Ofer_EMP} using successive schemes (where the target rate is the capacity of the respective modulu-additive channel). The two concepts in achieving the empirical rates differ in various factors such as complexity and the amount of feedback and randomization required. The successive schemes require less common randomness but assume perfect feedback, while the schemes based on rateless coding require less (asymptotically 0 rate) feedback but potentially more randomness.

As noted the technique we use here is similar to that of \cite{Eswaran} in its high level structure, while the structure of the rateless decoder is similar to \cite{Shulman}'s (chapter 3). The application of this scheme to individual inputs and outputs and the extension to real-valued models requires proof and especially issues such as abnormal behavior of specific (e.g. last) symbols have to be treated carefully. The result of \cite{Eswaran} cannot be applied directly to individual channels since the channel model cannot be extracted based on the input and output sequences alone, and in the later both the model and the sequence are assumed to be fixed (over common randomness).

\subsection{Statement of the main result}\label{sec:rate_adaptive_theorems}
In this section we prove the following theorems, relating to the definitions given in section \ref{sec:definitions}:

\begin{theorem}[Rate adaptive, discrete channels]\label{theorem:discrete_adaptive}
Given discrete input and output alphabets $\mathcal{X,Y}$, for every $P_e>0$, $P_A>0$, $\delta>0$ and prior $Q(x)$ over $\mathcal{X}$ there is $n$ large enough and random encoder and decoder with feedback and variable rate over block size $n$ with a subset $J \subset\mathcal{X}^n$, such that:
\begin{itemize}
\item The distribution of the input sequence is $\vr{x} \sim Q^n$ independently of the feedback and message
\item The probability of error is smaller than $P_e$ for any $\vr x, \vr y$
\item For any input sequence $\vr x \not\in J$ and output sequence $\vr{y} \in \mathcal{Y}^n$ the rate is
$R \geq \hat I (\vr{x},\vr{y}) - \delta$
\item The probability of $J$ is bounded by $\Pr(\vr x \in J) \leq P_A$
\end{itemize}
\end{theorem}

\begin{theorem}[Rate adaptive, continuous channels]\label{theorem:continuous_adaptive}
Given the channel $\mathbb{R} \to \mathbb{R}$ for every $P_e>0$, $P_A>0$, $\delta>0$, $\bar R>0$, and power $P>0$
there is $n$ large enough and random encoder and decoder with feedback and variable rate over block size $n$ with a
subset $J \subset\mathbb{R}^n$, such that
\begin{itemize}
\item The distribution of the input sequence is $\vr{x} \sim \Normal(0,P)^n$ independently of the feedback and message
\item The probability of error is smaller than $P_e$ for any $\vr x, \vr y$
\item For any input sequence $\vr x \not\in J$ and output sequence $\vr{y} \in \mathbb{R}^n$ the rate is
$R \geq \min \left[ \half \log \left( \frac{1}{1-\hat\rho(\vr x, \vr y)^2} \right) - \delta, \bar R \right] $
\item The probability of $J$ is bounded by $\Pr(\vr x \in J) \leq P_A$
\end{itemize}
\end{theorem}

\begin{figure}
\center
  \includegraphics[width=8cm]{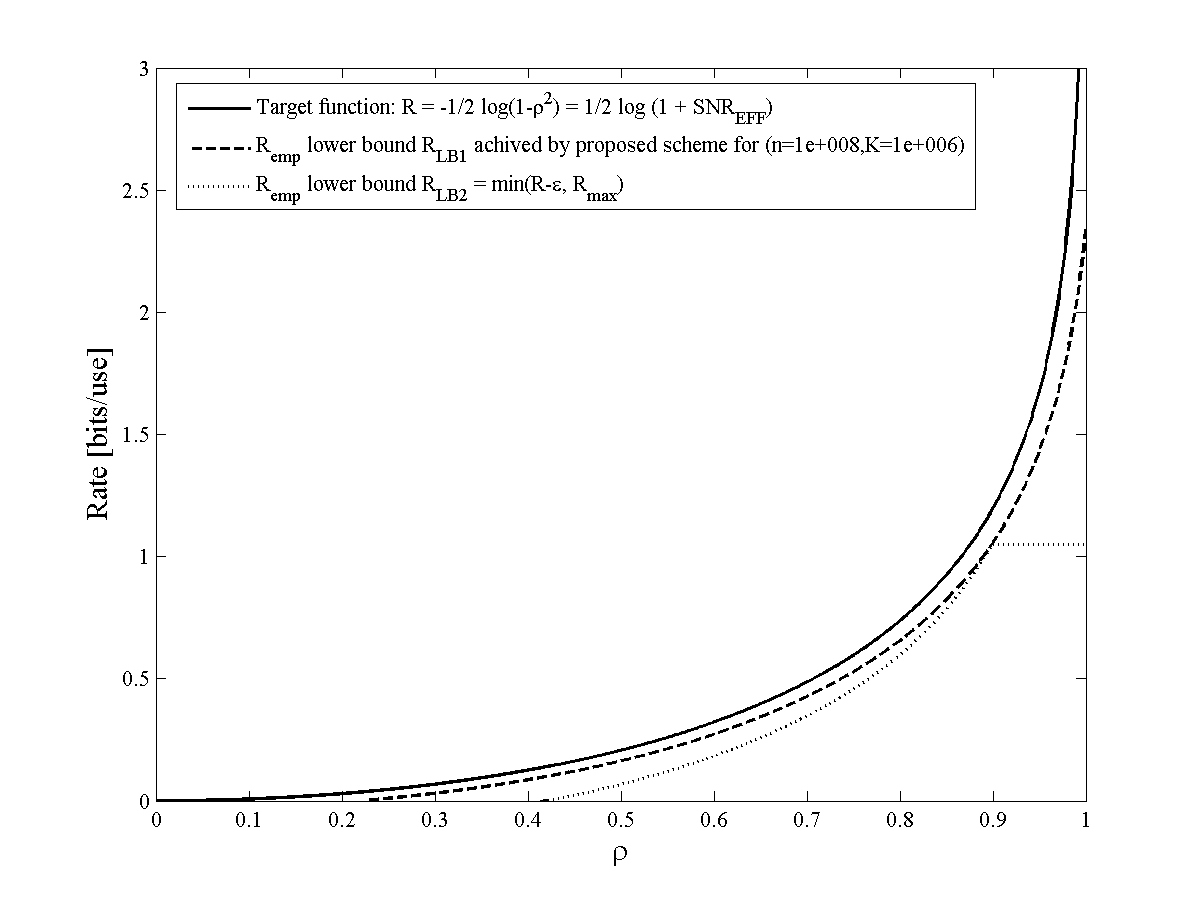}
  \includegraphics[width=8cm]{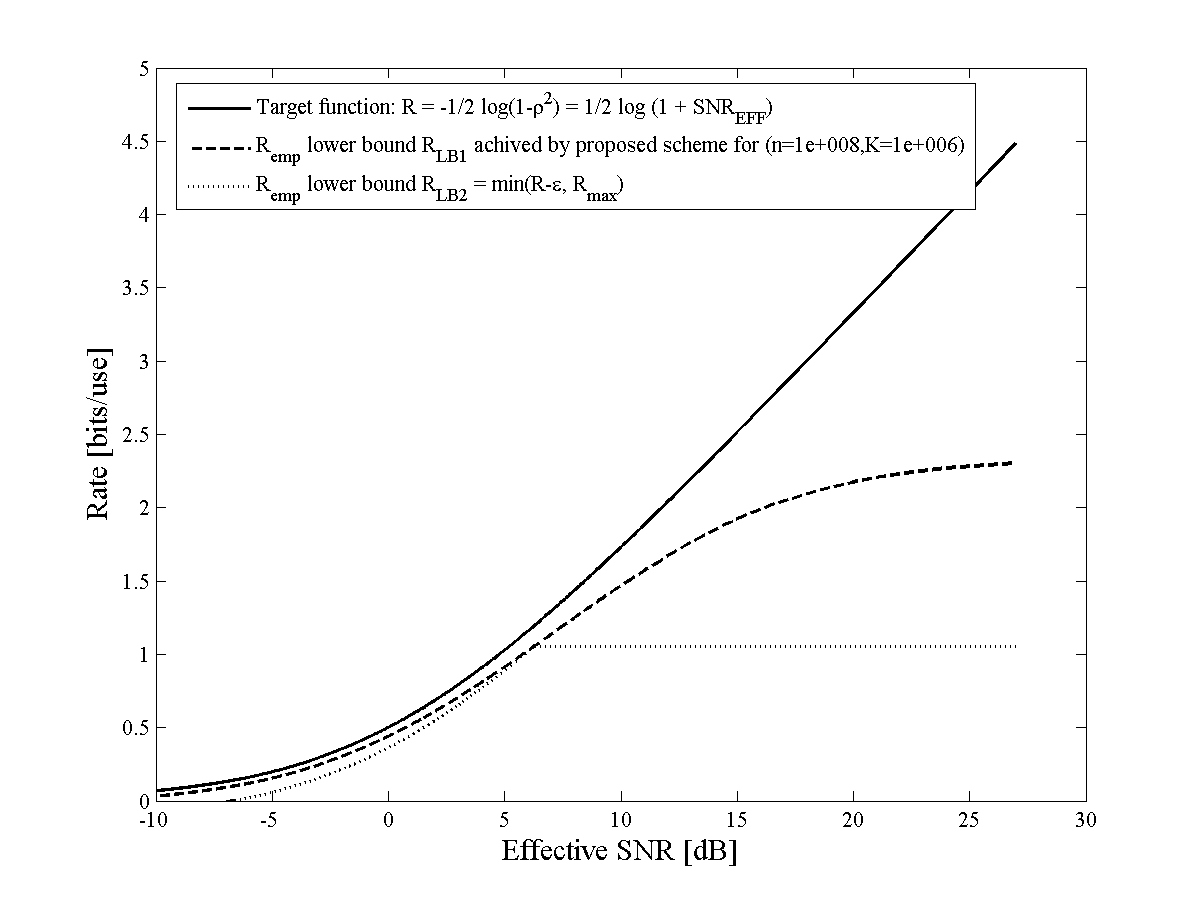}
  \caption{Illustration of $\Remp$ lower bound of theorem \ref{theorem:continuous_adaptive} ($\RLBTWO$) and the lower bound $\RLBONE$ shown in the proof in section \ref{sec:rate_analysis_continuous}, as a function of
$\rho$ (top) and the effective SNR $= \frac{\rho^2}{1-\rho^2}$ (bottom). Parameters appear in table \ref{table:rate_in_continuous_theorem_params} in the appendix}\label{fig:rate_in_continuous_theorem}
\end{figure}

Note that in the last theorem we do not have uniform convergence of the rate function in $\vr x, \vr y$. Unfortunately our scheme is limited by having a maximum rate for each $n$, and although the maximum rate tends to infinity as $n \to \infty$, we cannot guarantee uniform convergence for each $n$ in the continuous case, where the target rate may be unbounded. The rates in the theorems are the minimal rates, and in certain conditions (e.g. a channel varying in time) higher rates may be achieved by the scheme proposed below.

Regarding the set $J$ as we shall see in the sequel there are some sequences for which poor rate is obtained, and since we committed to an input distribution we cannot avoid them (one example is the sequence of $\half n$ zeros followed by $\half n$ ones, in which at most one block will be sent). However there is an important distinction between claiming for example that "for each $\vr y$ the probability of $R < \Remp$ is at most $P_A$" and the claim made
in the theorems that "$R < \Remp$ only when $\vr x$ belongs to a subset $J$ with probability at most $P_A$". The first claim is weaker since smartly chosen $\vr y$ may increase the probability (see figure \ref{fig:Bad_x_sequences}). This is avoided in the second claim. A consequence of this definition is that the probability of $R < \Remp$ is bounded by $P_A$ for any conditional probability $\Pr(\vr y | \vr x)$ on the sequences. This issue is further discussed in section \ref{sec:rate_adaptive_perliminaries}.

Note that the probability $P_A$ could be absorbed into $P_e$ by a simple trick, but this seems to make the Theorem less insightful. After reception the receiver knows the input sequence in probability of at least $1-P_e$ and may calculate the empirical mutual information $\hat I(\vr x, \vr y)$. If the rate achieved by the scheme we will describe later falls short of $\hat I(\vr x, \vr y)$ it may declare a rate of $R=\hat I(\vr x, \vr y)$ (which will most likely result in a decoding error). This way the receiver will never declare a rate which is lower than $\hat I(\vr x, \vr y)$ unless there is an error, and we could avoid the restriction $\vr x \not\in J$ required for achieving $\Remp$, but on the other hand, the error probability becomes conditioned on the set $J$. The question whether the set $J$ itself is truly necessary (i.e. is it possible to attain the above Theorems with $J=\emptyset$) is still open.

Figure (\ref{fig:rate_in_continuous_theorem}) illustrates the lower bound for $\Remp$ presented by Theorem \ref{theorem:continuous_adaptive} ($\RLBTWO$) as well as a (higher) lower bound $\RLBONE$ for the rate achieved by the proposed scheme (see section \ref{sec:rate_analysis_continuous}, Eq.(\ref{eq:cont_rate_analysis9})). The parameters generating these curves appear in table \ref{table:rate_in_continuous_theorem_params} in the appendix.

We prove the two theorems together. First we define the scheme, and in the next section we analyze its error performance and rate and show it achieves the promise of the theorems. Throughout this section and the following one we use $n$ to denote the length of a complete transmission, and $m$ to denote the length of a single block.

\subsection{A proposed rate adaptive scheme}\label{sec:rate_adaptive_scheme}
The following communication scheme sends $B$ indices from $\{1,\ldots,M\}$ over $n$ channel uses (or equivalently sends the number $\theta \in [0,1)$ in resolution $M^{-B}$), where $M$ is fixed, and $B$ varies according to empirical channel behavior. The building block is a rateless transmission of one of $M$ codewords ($K \equiv \log(M)$ information units), which is iterated until the $n$-th symbol is reached.

The transmit distribution $Q$ is an arbitrary distribution for the discrete case and $Q = \Normal(0,P)$ for the continuous case. We define the decoding metric as the empirical rate:
\begin{equation}\label{eq:def_mu}
\Remp(\vr x, \vr y) \equiv  \left\{ \begin{array}{ll}
\hat I (\vr x, \vr y) & \textrm{discrete}\\
\half \log \left( \frac{1}{1 - \hat\rho^2(\vr x, \vr y)} \right) & \textrm{continuous}\end{array} \right.
\end{equation}
The codebook $C_{M \times n}$ consists of $M$ codewords of length $n$, where all $M \times n$ symbols are drawn i.i.d. $\sim Q$ and known to the sender and receiver. For brevity of notation we denote $\Remp^m(\vr x, \vr y)$ instead of $\Remp(\vr x_1^m, \vr y_1^m)$. $k$ denotes the absolute time index $1 \leq k \leq n$. Block $b$ starts from index $k_b$, where $k_1=1$. $m = k - k_b + 1$ denotes the time index inside the current block.

In each rateless block $b=1,2,\ldots$, a new index $i = i_b \in \{1,\ldots,M\}$ is sent to the receiver using the following procedure:
\begin{enumerate}
\item The encoder sends index $i$ by sending the symbols of codeword $i$:
\begin{equation}x_k = C_{i,k}\end{equation}
Note that different blocks use different symbols from the codebook.
\item The encoder keeps sending symbols and incrementing $k$ until the decoder announces the end of the block through the feedback link.
\item The decoder announces the end of the block after symbol $m$ in the block if for any codeword $x_i$ :
\begin{equation}\label{eq:termination_condition}
\Remp^m(\vr x_i, \vr y) \equiv \Remp \left( (\vr x_i)_{k_b}^k, \vr y_{k_b}^k \right) \geq \mu^*_m
\end{equation}
where $\mu^*_m$ is a fixed threshold per symbol defined in Eq.(\ref{eq:termination_threshold}) below.
\item When the end of block is announced one of the $i$ fulfilling Eq.(\ref{eq:termination_condition}) is determined as the index of the decoded codeword $\hat i_b$ (breaking ties arbitrarily).
\item Otherwise the transmission continues, until the $n$-th symbol is reached. If symbol $n$ is reached without fulfilling Eq.(\ref{eq:termination_condition}), then the last block is terminated without decoding.
\end{enumerate}

After a block ends, $b$ is incremented and if $k < n$ a new block starts at symbol $k_b = k+1$. After symbol $n$ is reached the transmission stops and the number of blocks sent is $B=b-1$.

The threshold $\mu^*_m$ is defined as:
\begin{multline}\label{eq:termination_threshold}
\mu^*_m = \frac{K}{m-s} + \frac{1}{m-s}\log \left( \frac{n}{P_e} \right) + \delta_m
=\\=
 \left\{ \begin{array}{ll}
\frac{K + \log \left( \frac{n}{ P_e} \right) + |\mathcal{X}||\mathcal{Y}| \log(m+1)}{m}  & \textrm{discrete}\\
\frac{K + \log \left( \frac{2n}{P_e} \right)}{m-1} & \textrm{continuous}\end{array} \right.
\end{multline}
where $s=0$ for the discrete case and $1$ for the continuous case and $\delta_m$ is defined in Lemma \ref{lemma:pairwise_discrete} for the discrete case and equals $\frac{\log(2)}{m-1}$ for the continuous case. The threshold $\mu^*_m$ is tailored to achieve the designated error probability and is composed of 3 parts. The first part requires that the empirical rate $\Remp$ would approximately equal the transmission rate of the block $\frac{K}{m}$, which guarantees there is approximately enough mutual information to send $K$ information units. The second part is an offset responsible for guaranteeing error probability bounded by $P_e$ over all the blocks in the transmission. The third part $\delta_m$ compensates the overhead terms in Lemmas \ref{lemma:pairwise_discrete},\ref{lemma:pairwise_continuous}.

The scheme achieves the claims of Theorems \ref{theorem:discrete_adaptive},\ref{theorem:continuous_adaptive}
with a proper choice of the parameters (discussed in section \ref{sec:rate_analysis}). Note that the scheme uses feedback rate of $1$ bit/use however it is easy to show any positive feedback rate is sufficient (see section \ref{sec:rate_analysis}), therefore we can claim the theorems hold with "zero rate" feedback.

We devote the next section to the analysis of the error probability and rate of the scheme, showing it attains Theorems \ref{theorem:discrete_adaptive},\ref{theorem:continuous_adaptive}. Unfortunately although the scheme is simple, the current analysis we have is somewhat cumbersome.

\section{Proof of the main result}\label{sec:analysis}
In this section we analyze the adaptive rate scheme presented and show it achieves Theorems \ref{theorem:discrete_adaptive},\ref{theorem:continuous_adaptive}. Before analyzing the scheme we develop some general results pertaining to the convexity of the mutual information and correlation factors over sub-vectors. The proof of the error probability is simple (based on the construction of $\mu^*_m$) and common to the two cases. The proof of the achieved rate is more complex and performed separately for each case.

\subsection{Preliminaries}\label{sec:rate_adaptive_perliminaries}

\subsubsection{Likely convexity of the mutual information}\label{sec:likely_convexity_of_MI}
A property which would be useful for the analysis is $\cup$-convexity of the empirical mutual information with respect to joint empirical distributions $\hat P_{(\vr x, \vr y)}(x,y)$ measured over different sub-vectors, so for example we would like to have for $0 \leq m \leq n$:
\begin{equation}\label{eq:erronous_convexity}
\hat I(\vr x_1^n; \vr y_1^n) \leq \frac{m}{n} \cdot \hat I(\vr x_1^m; \vr y_1^m) + \left(1-\frac{m}{n}\right) \cdot \hat I(\vr x_{m+1}^n; \vr y_{m+1}^n)
\end{equation}
which would guarantee that if we achieve a rate equal to the empirical mutual information over the two sections $0 \leq k \leq m$ and $m < k \leq n$, then we would achieve the empirical mutual information over the entire vector $0 \leq k \leq n$. However this property does not hold in general since the mutual information is not convex with respect to the joint distribution. The mutual information $I(P,W)$ is known to be convex $\cup$ with respect to $W$ and concave $\cap$ with respect to $P$, so if, for example, the conditional distributions over the sections $[1,m]$ and $[m+1, n]$ are equal and only the distribution of $\vr x$ differs, the condition would in general not hold. On the other hand should the empirical distributions of $\vr x_1^{m}$ and $\vr x_{m+1}^n$ be equal, then the empirical mutual information expressions appearing in Eq.(\ref{eq:erronous_convexity}) would differ only in the conditional distributions of $\vr y$ w.r.t $\vr x$ and the assertion would hold. Since we generate $\vr x$ by i.i.d. drawing of its elements the empirical distributions converge to the prior $Q$, and we would expect that if the size of both regions $m$ and $m-n$ is large enough the convexity would hold up to a fraction $\epsilon$ in high probability. We show below that such convexity holds under even milder conditions. The cases in which this approximate convexity is used later on can serve as examples of the difference between the individual model used here and probabilistic models (including the individual noise sequence). We use the lemma to:
\begin{enumerate}
\item Bound the loss due to insufficient utilization of the last symbol in each rateless encoding block.
\item Bound the loss due to not completing the last rateless encoding block.
\item Show that the average rate (empirical mutual information) over multiple blocks equals at least the mutual information measured over the blocks together
\end{enumerate}
Had the rate been averaged over multiple sequences $\vr x$ rather than obtained for a specific sequence, the regular convexity of the mutual information with respect to the channel distribution would have been sufficient. The property is formalized in the following lemma:

\begin{lemma}[Likely convexity of mutual information]\label{lemma:likely_convexity_of_MI}
Let $\{A_i\}_{i=1}^p$ define a disjoint partitioning of the index set $\{1, \ldots, n\}$, i.e. $\bigcup_i A_i = \{1, \ldots, n\}$ and $A_i \cap A_j = \emptyset \text{ for } i \neq j$. $\vr x$ , $\vr y$ are n-length sequences, and $\vr x_A$, $\vr y_A$ define the sub-sequences of $\vr x$, $\vr y$ (resp.) over the index set $A$. Let the elements of $\vr x$ be chosen i.i.d. with distribution $Q$. Then for any $\Delta>0$ there is a subset $J_{\Delta} \subset \mathcal{X}^n$
such that: \begin{equation}
\forall \vr x \not\in J_{\Delta}, \vr y \in \mathcal{Y}^n : \hspace{3ex}
\sum_{i=1}^p  \frac{|A_i|}{n} \hat I (\vr x_{A_i}; \vr y_{A_i}) \geq  \hat I (\vr x; \vr y)-\Delta
\end{equation}
And
\begin{equation}
Q^n \left\{J_{\Delta} \right\} \leq \exp \left( -n \left(\Delta - \tilde{\delta}_n \right) \right)
\end{equation}
With $\tilde{\delta}_n = p |\mathcal{X}| \cdot \frac{\log(n+1)}{n} \to 0$.
\end{lemma}
The lemma does not claim that convexity holds with high probability, but rather that any positive deviation from convexity may happen only on a subset of $\vr x$ with vanishing probability.
It is surprising that the bound does not depend on $\vr y$, $Q$ and the size of the subsets, and only weakly depends on the number of subsets.

\begin{figure}
  \center
  \includegraphics[width=8cm]{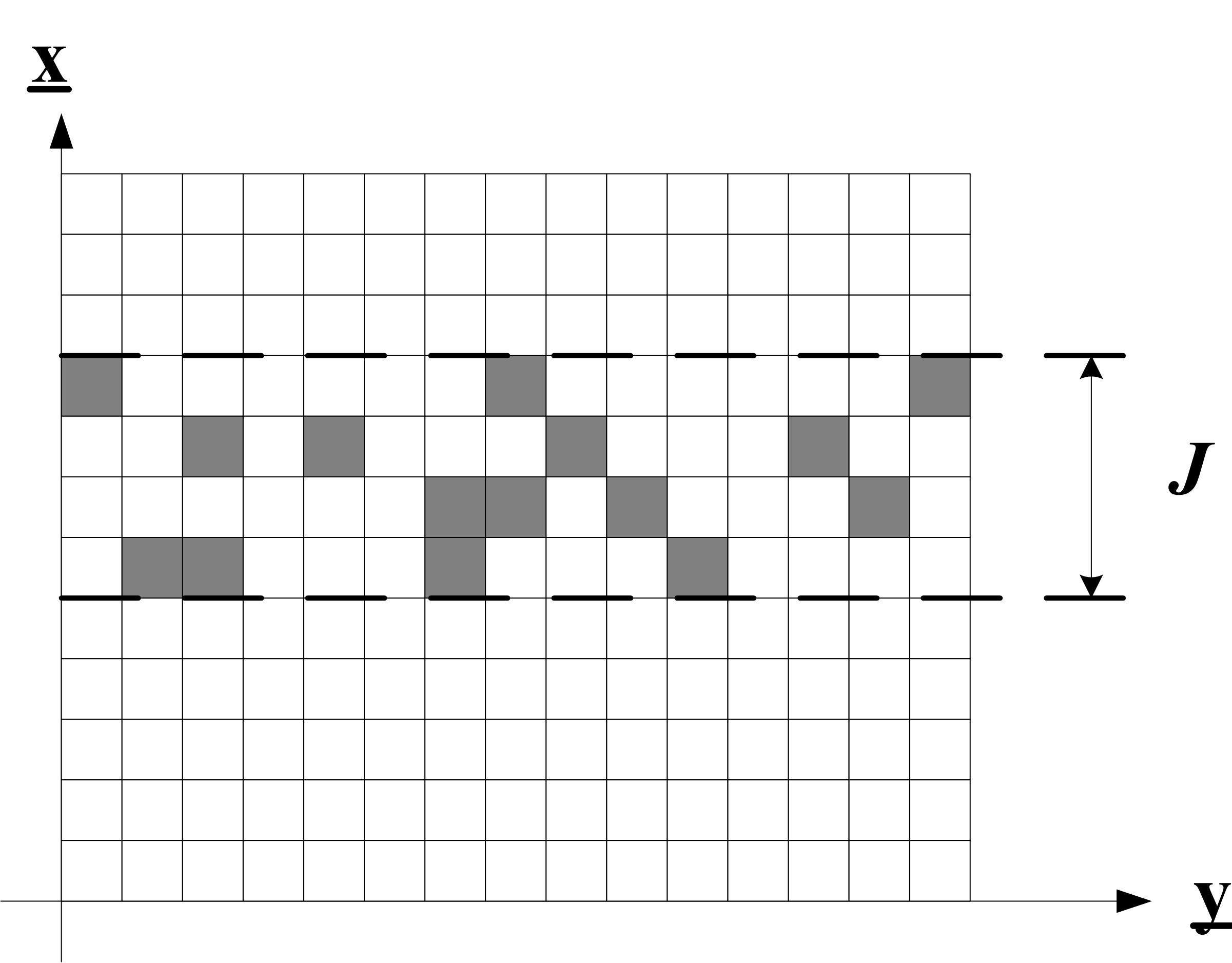}\\
  \caption{Illustration of bad sequences and lemma \ref{lemma:likely_convexity_of_MI}}\label{fig:Bad_x_sequences}
\end{figure}

Before proving the lemma we emphasize a delicate point: the lemma does not only claim that for each $\vr y$ the probability of deviation from convexity is small, but makes a stronger claim that apart from a subset of the $\vr x$ sequences with vanishing probability, convexity always holds independently of $\vr y$. This distinction is important since this lemma defines a set of "bad" input sequences that fail our scheme. In these sequences there exists a partitioning that yields an excessive deviation from the distribution $Q$ between rateless blocks. As an example of such a sequence consider the binary channel and the input sequence $0^{n/2}1^{n/2}$ ($n/2$ zeros followed by $n/2$ ones). This sequence is bad since it guarantees that on one hand at most one block will be received (since at most one block includes both $0$-s and $1$-s at the input), but on the other hand the zero order empirical input distribution is good ($Ber(\half)$), so potentially we have the combination of high empirical mutual information with low communication rate. The sequences that deviate from convexity are a function of the output $\vr y$. Had we only
bounded the probability of deviation from convexity to occur for each $\vr y$ individually, then a potential adversary
could have increased this probability by determining $\vr y$ (given $\vr x$) such that $\vr x$ will be a bad sequence
with respect to this $\vr y$. To avoid this, we claim that there is a fixed group of $\vr x$ such that if the
sequence is not in the group, approximate convexity holds regardless of $\vr y$. This is illustrated
in fig.(\ref{fig:Bad_x_sequences}) where the dark spots mark the pairs $(\vr x, \vr y)$ for which convexity does not hold.

\textit{Proof of lemma \ref{lemma:likely_convexity_of_MI}}:
Define the vector $\vr u$ denoting the subset number of each element $\vr u_k = i \hspace{1em} \forall k \in A_i$. Then $\hat I (\vr x_{A_i}; \vr y_{A_i}) = \hat I (\vr x, \vr y | \vr u = i)$, and $\hat P_{\vr u}(i) = \frac{|A_i|}{n}$, therefore we can write the weighted sum of empirical mutual information over the partitions, as a conditional empirical mutual information:
\begin{multline}
\sum_{i=1}^p \left( \frac{|A_i|}{n} \hat I (\vr x_{A_i}; \vr y_{A_i}) \right) =
\sum_{i=1}^p \hat P_{\vr u}(i) \hat I (\vr x; \vr y | \vr u = i)
=\\=  \hat I(\vr x; \vr y | \vr u)
\end{multline}
Using the chain rule for mutual information (see \cite{Cover} section 2.5):
\begin{multline}
\hat I (\vr x; \vr y) - \hat I(\vr x; \vr y | \vr u)
=  \hat I (\vr x; \vr y) - \left(\hat I(\vr x; \vr y \vr u) -  \hat I(\vr x; \vr u)\right)
=\\=
\hat I(\vr x; \vr u) - \hat I(\vr x; \vr u | \vr y)  \leq  \hat I(\vr x; \vr u)
\end{multline}
Define the set $J_{\Delta} = \{\vr x : \hat I(\vr x; \vr u) > \Delta \}$, then
\begin{equation}
\forall \vr x \not\in J_{\Delta}, \vr y: \hat I (\vr x; \vr y) - \hat I(\vr x; \vr y | \vr u) \leq \hat I(\vr x; \vr u) \leq \Delta
\end{equation}
And since $\vr x$ is chosen iid and $\vr u$ is a fixed vector, we have from Lemma \ref{lemma:pairwise_discrete}:
\begin{equation}
\Pr \left( \vr x \in J_{\Delta} \right) \leq \exp \left( -n \left(\Delta - \tilde{\delta}_n \right) \right)
\end{equation}
with $\tilde{\delta}_n = |\mathcal{X}||\{1,\ldots,p\}|\frac{\log(n+1)}{n}$.
\myendofproof

Note that if the distribution of $\vr x$ is the same over all partitions then $\hat H(\vr x| \vr u) = \hat H(\vr x)$ therefore $\hat I(\vr x; \vr u) = 0$ and the empirical mutual information will be truly convex.

\subsubsection{Likely convexity of the correlation factor}\label{sec:likely_convexity_of_rho}
For the continuous case we use the following property which somewhat parallels Lemma \ref{lemma:likely_convexity_of_MI}. The reasons for not following the same path as the discrete case will be explained in the sequel (subsection \ref{sec:rate_analysis}). Unfortunately the proof is very technical and less elegant and will therefore be expelled to the appendix (appendix-\ref{appendix:likely_convexity_of_rho}). Note that again the bound does not depend on the size of the subsets.
\begin{lemma}[Likely convexity of $\hat\rho^2$]\label{lemma:likely_convexity_of_rho}
Define $\{A_i\}_{i=1}^p$ as in Lemma \ref{lemma:likely_convexity_of_MI}. Let $\vr x$ , $\vr y$ be $n$-length sequences and define the correlation factors of the sub-sequences, and the overall correlation factor as
\begin{equation}
\hat\rho_i = \frac{\lvert \vr x_{A_i}^T \vr y_{A_i} \rvert}{\lVert \vr x_{A_i} \rVert \cdot \lVert \vr y_{A_i} \rVert} \hspace{2em} \text{and} \hspace{2em} \hat\rho = \frac{\lvert \vr x^T \vr y \rvert}{\lVert \vr x \rVert \cdot \lVert \vr y \rVert}
\end{equation}
respectively. Let $\vr x$ be drawn i.i.d from a Gaussian distribution $\vr x \sim \Normal(0,P)$. Then for any $0<\Delta\leq \frac{1}{7}$ there is a
subset $J_{\Delta} \subset \mathbb{R}^n$ such that:
\begin{equation}
\forall \vr x \not\in J_{\Delta}, \vr y \in \mathbb{R}^n : \hspace{3ex}
\sum_{i=1}^p  \frac{|A_i|}{n} \hat\rho_i^2 \geq \hat\rho^2 - \Delta
\end{equation}
And
\begin{equation}
\Pr \left\{\vr x \in J_{\Delta} \right\} \leq 2^p e^{ - n \Delta^2/8}
\end{equation}
I.e. there is a subset with high probability on which the mean of the correlation factors does not fall considerably below the overall correlation factor.
\end{lemma}

\subsubsection{Likely convexity with dependencies}\label{sec:likely_convexity_with_dependency}
The properties of likely convexity defined in the previous sections pertain to a case where the partition of the $n$ block is fixed and $\vr x$ is drawn i.i.d. However in the transmission scheme we described, the partition varies in a way that depends on the value of $\vr x$ (through the decoding decisions and the empirical mutual information), which may, in general, change the probability of the convexity property with a given $\Delta$ to occur. Although it stands to reason that the variability of the block sizes in the decoding process reduces the probability to deviate from convexity since it tends to equalize the amount of mutual information in each rateless block, for the analysis we assume an arbitrary dependence, and assume that the size of the set $J$ increases by factor of the number of
possible partitions, as explained below.

Denote a partition by $\pi = \{A_i\}_{i=1}^p$  (as defined in Lemmas \ref{lemma:likely_convexity_of_MI},\ref{lemma:likely_convexity_of_rho}) and the group of all possible partitions (for a given encoder-decoder) by $\Pi$. For each partition $\pi$  from Lemmas \ref{lemma:likely_convexity_of_MI},\ref{lemma:likely_convexity_of_rho}
there is a subset $J(\pi)$ with probability bounded
by $p_J$ outside which approximate convexity (as defined in the lemmas) holds. Then approximate convexity is guaranteed to hold for
 $\vr x \not\in J \equiv \displaystyle \bigcup_{\pi \in \Pi} J(\pi)$, where the probability of the set $J$ is bounded by the
 union bound:
 \begin{equation}
\Pr(\vr x \in J) = \Pr \left( \bigcup_{\pi \in \Pi} (\vr x \in J(\pi))  \right) \leq |\Pi| \cdot p_J
\end{equation}


Now we bound the number of partitions. In the two cases we will deal with in section \ref{sec:rate_analysis} the number
of subsets can be bounded by $p_{\max}$, and all subsets but one contain continuous indices.
Therefore the partition is completely defined by the start and end indices of $p_{\max}-1$ subsets (allowed to overlap if there are less than $p_{\max}$ subsets), thus $|\Pi| \leq n^{2p_{\max}-2} < n^{2p_{\max}}$ and we have
\begin{equation}\label{eq:likely_convexity_with_dependency}
\Pr (J) \leq n^{2 p_{\max}} \cdot p_J = \exp( 2 p_{\max} \log(n) ) \cdot p_J
\end{equation}
where $p_J$ is defined in the previous lemmas. So for our purposes we may say that these lemmas hold even if the
partition depends on $\vr x$ with an appropriate change in the probability of $J$.

\begin{table*}[t]
\caption{Summary of definitions and references for the discrete and continuous cases}\label{table:summary_of_definitions}
\center
\begin{tabular}{|p{5cm}|p{5cm}|p{5cm}|}
\hline
Item            &   Discrete case               &   Continuous case \\ \hline
Input distribution &    Any $Q$               &   $Q = \Normal(0,P)$ \\ \hline
Decoding metric &   $\Remp(\vr x, \vr y) \equiv \hat I (\vr x, \vr y)$    & $\Remp(\vr x, \vr y) \equiv  \half \log \left( \frac{1}{1 - \hat\rho^2(\vr x, \vr y)} \right)$ \\ \hline
Decoder &   maximize $\Remp(\vr x, \vr y)$ $\Leftrightarrow$ maximize $\hat I (\vr x, \vr y)$    & maximize $\Remp(\vr x, \vr y)$ $\Leftrightarrow$ maximize $|\hat \rho(\vr x, \vr y)|$ \\ \hline
Pairwise error probability $\Pr(\Remp \geq t)$ & $\leq \exp(-n (t - \delta_n))$ (Lemma \ref{lemma:pairwise_discrete}) & $\leq 2 \exp(-(n-1) t)$ (Lemma \ref{lemma:pairwise_continuous}) \\ \hline
Likely convexity condition ($\forall \vr x \not\in J_{\Delta}, \vr y \in \mathcal{Y}^n$ with $\lambda_i \equiv \frac{1}{n}|A_i|$)
    & $\sum_{i=1}^p  \lambda_i \hat I (\vr x_i; \vr y_i) \geq  \hat I (\vr x; \vr y)-\Delta$ (Lemma \ref{lemma:likely_convexity_of_MI})
    & $\sum_{i=1}^p  \lambda_i \hat\rho_i^2 \geq \hat\rho^2 - \Delta $ (Lemma \ref{lemma:likely_convexity_of_rho}) \\ \hline
Likely convexity probability ($\Pr(\vr x \not\in J_{\Delta})$, fixed partitioning) & $\geq 1 - \exp \left( -n \left(\Delta - \tilde{\delta}_n \right) \right)$ & $\geq 1 - 2^p e^{ - n \Delta^2/8}$ \\ \hline
\end{tabular}
\end{table*}

\subsection{Error probability analysis}\label{sec:error_analysis}
In this subsection we show the probability to decode incorrectly any of the $B$ indices is smaller than $P_e$.

With $\Remp$ defined in Eq.(\ref{eq:def_mu}), we have from Lemma \ref{lemma:pairwise_continuous} that under the conditions of the lemma $\Pr(\Remp \geq t) = \Pr(|\hat\rho| \geq R_2^{-1}(t)) \leq 2 \exp(-(n-1) t)$. Then combining Lemmas \ref{lemma:pairwise_discrete} and \ref{lemma:pairwise_continuous}, we may say that for any $\vr y_1^m$ the probability of $\vr x_1^m$ generated i.i.d. from the relevant prior to have $\Remp \geq t$ is bounded by:
\begin{equation}\label{eq:unified_pairwise}
Q^m(\Remp(\vr x_1^m, \vr y_1^m) \geq t) \leq \exp \left( -(m-s) (t - \delta_m) \right)
\end{equation}
where
\begin{equation}\label{eq:unified_pairwise_delta}
\delta_m = \left\{ \begin{array}{ll}
|\mathcal{X}||\mathcal{Y}|\frac{\log(m+1)}{m} & \textrm{discrete}\\
\frac{\log 2}{m-1} & \textrm{continuous}\end{array} \right.
\end{equation}
And
\begin{equation} s = \left\{ \begin{array}{ll}
0 & \textrm{discrete}\\
1 & \textrm{continuous}\end{array} \right.
\end{equation}

An error might occur if at any symbol $1 \leq k \leq n$ an incorrect codeword meets the termination condition Eq.(\ref{eq:termination_threshold}). The probability that codeword $j \neq i$ meets Eq.(\ref{eq:termination_threshold}) at a specific symbol $k$ which is the $m$-th symbol of a rateless block is bounded by:
\begin{multline}
\Pr(\Remp^m(\vr x_j, \vr y) \geq \mu^*_m) \leq \exp \left( -(m-s) (\mu^*_m - \delta_m) \right)
=\\= \exp \left( -\left[ K + \log \left( \frac{n}{P_e} \right) \right] \right) = \frac{P_e}{n \exp(K)} = \frac{P_e}{M n}
\end{multline}

The probability of any erroneous codeword to meet the threshold at any symbol is bounded by the union bound:
\begin{multline}
\Pr(\textrm{error}) \leq \Pr \left\{ \bigcup_{k=1}^n \bigcup_{j \neq i} \left(\mu_m(\vr x_j, \vr y)\geq \mu^*_m \right) \right\}
\leq \\ \leq
n (M-1) \frac{P_e}{M n} < P_e
\end{multline}

The first inequality is since the correct codeword might be decoded even if an erroneous codeword met the threshold. Although the index $m$ in the expression above depends on $k$ and the specific sequences $\vr x, \vr y$ in an unspecified way, the assertion is true since the probability of the event in the union has an upper bound independent of $m$.

\subsection{Rate analysis}\label{sec:rate_analysis}
Roughly speaking, since $\mu^*_m \approx \frac{K}{m}$, if no error occurs, the correct codeword crossed the threshold when $\Remp^m(\vr x_i, \vr y) \approx \frac{K}{m}$ therefore the rate achieved over a rateless block is $R_b = \frac{K}{m} \approx \Remp^m(\vr x_i, \vr y)$, and due to the approximate convexity by achieving the above rate on each block separately we meet or exceed the rate $\Remp(\vr x, \vr y)$ over the entire transmission. However in a detailed analysis we have the following sources of rate loss:
\begin{enumerate}
\item The offsets inserted in $\mu^*_m$ to meet the desired error probability
\item The offset from convexity (Lemma \ref{lemma:likely_convexity_of_MI}) introduced by the slight differences in empirical distribution of $\vr x$ between the blocks
\item Unused symbols:
 \begin{enumerate}
 \item The last symbol of each block is not fully utilized, as explained below
 \item The last (unfinished) block is not utilized
 \end{enumerate}
\end{enumerate}

Regarding the last symbol of each block, note that after receiving the previous symbol the empirical mutual information is below the threshold, and at the last symbol it meets or exceeds the threshold. However the proposed scheme does not gain additional rate from the difference between the mutual information and the threshold, and thus it loses with respect to its target (the mutual information over the block) when this difference is large. Here a "good" channel works adversely to our worse. Since we operate under an individual channel regime, the increase of the mutual information at the last symbol is not bounded to the average information contents of a single symbol. This is especially evident in the continuous case where the empirical mutual information is unbounded. A high value of $y$ together with high value of $x$ at the last symbol causes an unbounded increase in $\Remp$: if we choose $\vr x_m,\vr y_m \to \infty$ then $\rho \to 1$ regardless of the history $\vr x_1^{m-1}, \vr y_1^{m-1}$. Therefore over a single block we might have an arbitrarily low rate ($|\hat\rho|$ is small over the $m-1$ first symbols) and arbitrarily large $\Remp$. In the discrete case this phenomenon exists but is less accented (consider for example the sequences $\vr x = \vr y = 0^{n-1}1 = (0,\ldots,0,1)$)
Similarly regarding the last block, the fact that the length of the block may be bounded does not mean the increase in the empirical mutual information can be bounded as well. We use the approximate convexity (Lemma \ref{lemma:likely_convexity_of_MI}) to show the last two losses are bounded for most $\vr x$ sequences.

Note that by the same argument that shows the loss from not utilizing the last symbol vanishes asymptotically, it is easy to show that feeding back the block success information only once every $1/\epsilon$ symbols
thereby decreasing the feedback rate to $\epsilon$ does not decrease the asymptotical rate, since this is equivalent
to having $1/\epsilon$ unused symbols instead of one. Hence the scheme can be modified to operate with "zero rate" feedback. Similarly the scheme can operate with a noisy feedback channel by introducing in the feedback link a delay suitable to convey the decoder decisions with sufficiently low error rate over the noisy channel.

In addition to having rate losses the scheme also has a minimal rate and a maximal rate for each block length. The minimal rate is $\frac{K}{n}$ resulting from sending a single block. If channel conditions are worse ($\Remp < \frac{K}{n}$), no information will be sent. A maximal rate exists since at best $K$ information units could be sent every $2$ symbols (since for the continuous case $\mu^*_1 = \infty$ and for the discrete case $\Remp^1(x,y) = 0$ thus the decoding never terminates at the first symbol of the block), hence the maximum rate is $\frac{K}{2}$. As $n \to \infty$ we increase $K$ so that the minimum rate (and the rate offsets) tend to 0 and the maximum rate tends to $\infty$. The maximum rate is the reason that the scheme cannot approach the target rate $\Remp(\vr x_i, \vr y)$ uniformly in $\vr x,\vr y$ in the continuous case, since for some pairs of sequences the target rate (which is unbounded) may be much higher than the maximum rate. The rate $\bar R$ that we achieve in the proof of Theorem \ref{theorem:continuous_adaptive} is much smaller than the absolute maximum $\frac{K}{2}$. Note that successive schemes (such as Schalkwijk's \cite{Schalkwijk}) do not suffer from the problem of maximum rate.  For the discrete case the target rate is bounded by $\max (|\mathcal{X}|,|\mathcal{Y}|)$ therefore for sufficiently large $n$ the maximal rate $\frac{K}{2}$ exceeds $\max (|\mathcal{X}|,|\mathcal{Y}|)$ and we are able to show uniform convergence.

Although our target is the empirical mutual information over the $n$-block, an artifact of the partitioning to smaller blocks is that higher rates can be attained when the empirical conditional channel distribution varies over time, since by the convexity of mutual information with respect to the channel law the convex sum of mutual information over blocks exceeds the overall mutual information if these are not constant.

We now turn to prove the achieved rate. The total amount of information sent (with or without error) is $B \cdot K$ therefore the actual rate is
\begin{equation}
R_{\mathrm{act}} = \frac{B K}{n}
\end{equation}
We now endeavor to show this rate is close to or better than the empirical mutual information in probability of at least $P_A$ over the sequences $\vr x$, regardless of $\vr y$ and of whether a decoding error occurred.

The following definition of index sets in $\{1,\ldots,n\}$ is used for both the discrete and the continuous cases: $U_b = \{k\}_{k=k_b}^{k_{b+1}-2}$ denotes the channel uses of block $b$ \emph{except} the last one, $L_0$ collects the last channel uses of all the blocks $L_0 = \{k_b-1: b>1\}$, and $U_{B+1}$ denotes the indices of the un-decoded (last) block $U_{B+1} = \{k\}_{k=k_{B+1}}^{n}$ (including its last symbol), and is an empty set if the last block is decoded. The sets $\{U_b\}_{b=1}^{B+1}, L_0$ are disjoint and their union is $\{1,\ldots,n\}$. We denote the length of each block not including the last symbol by $m_b \equiv |U_b|$. From this point on we split the discussion and we start with the discrete case which is simpler.

\subsubsection{Rate analysis for the discrete case}\label{sec:rate_analysis_discrete}
We write $\mu^*_m$ as $\mu^*_m = \frac{K+\Delta_m}{m} \leq \frac{K+\Delta_{\mu}}{m}$ with
\begin{multline}
\Delta_m = \log \left( \frac{n}{ P_e} \right) + m\delta_m
=\log \left( \frac{n}{ P_e} \right) + |\mathcal{X}||\mathcal{Y}| \log(m+1)
\leq \\ \leq
\log \left( \frac{n}{P_e} \right) + |\mathcal{X}||\mathcal{Y}| \log(n+1)  \equiv \Delta_{\mu}
\end{multline}

From Lemma \ref{lemma:likely_convexity_of_MI} and Eq.(\ref{eq:likely_convexity_with_dependency}) we have that the following equation:
\begin{equation}\label{eq:convexity_in_rate_analysis}
 \hat I (\vr x; \vr y) - \Delta \leq \sum_{b=1}^{B+1} \left( \frac{m_b}{n} \hat I (\vr x_{B_b}; \vr y_{B_b}) \right) +
 \frac{|L_0|}{n} \hat I (\vr x_{L_0}; \vr y_{L_0})
 \end{equation}
is satisfied when $\vr x$ is outside a set $J_{\Delta}$ with probability of at most $\exp \left( -n \left(\Delta - \tilde{\delta}_n\right) \right)$ where $\tilde{\delta}_n = (B+2) |\mathcal{X}| \cdot \frac{\log(n+1)}{n} + 2 B_{\max} \frac{\log(n)}{n}$. We shall find $B_{\max}$ later on. To make sure the probability of $J$ is less than $P_A$ we require $\exp \left( -n \left(\Delta - \tilde{\delta}_n \right) \right) \leq P_A $ therefore
\begin{multline}
\Delta \geq \tilde{\delta}_n - \frac{1}{n} \log \left( P_A \right)
=\\= (B+2) |\mathcal{X}| \cdot \frac{\log(n+1)}{n} + 2 B_{\max} \frac{\log(n)}{n} - \frac{1}{n} \log \left( P_A \right)
\end{multline}
and we choose
\begin{equation}\label{eq:rate_analysis_delta}
\Delta = (3 B_{\max}+2) |\mathcal{X}| \cdot \frac{\log(n+1)}{n} - \frac{1}{n} \log \left( P_A \right)
\end{equation}

We now bound each element of Eq.(\ref{eq:convexity_in_rate_analysis}). Consider block $b$ with $m_b+1$ symbols. At the last symbol before decoding (symbol $m_b \equiv |U_b|$) none of the codewords, including the correct one crosses the threshold $\mu_m^*$, therefore:
\begin{equation}\label{eq:rate_analysis2}
\mu^*_{m_b} = \frac{K+ \Delta_{m_b}}{m_b}  > \hat I (\vr x_{U_b}; \vr y_{U_b})
\end{equation}
Specifically for the unfinished block we have at symbol $n$:
\begin{equation}\label{eq:rate_analysis3}
\mu^*_{m_{B+1}} = \frac{K+ \Delta_{m_{B+1}}}{m_{B+1}} > \hat I (\vr x_{U_{B+1}}; \vr y_{U_{B+1}})
\end{equation}
The way to understand these bounds is as guarantee on the shortness of the blocks given sufficient mutual information.
On the other hand, at the end of each block \emph{including} the last symbol (symbols $(k_b, k_b+m_b)$), since one of the sequences was decoded we have:
\begin{multline}\label{eq:rate_analysis2a}
\mu^*_{m_b+1} = \frac{K+ \Delta_{m_b+1}}{m_b+1}
\leq \\ \leq \max_{i} \hat I \left( \left( \vr x_i \right)_{k_b}^{k_b+m_b}; \vr y_{k_b}^{k_b+m_b} \right)
\leq
\log \min (|\mathcal{X}|,|\mathcal{Y}|) \equiv h_0
\end{multline}
Which we can use to bound the number of blocks, since $m_b+1 \geq \frac{K}{h_0}$ therefore
\begin{equation}\label{eq:rate_analysis2b}
 B \leq \sum_{b=1}^B \left( \frac{h_0}{K}(m_b+1) \right) \leq \frac{h_0 \cdot n}{K} \equiv B_{\max}
\end{equation}
As for the unused last-symbols we bound:
\begin{equation}\label{eq:rate_analysis4}
\hat I (\vr x_{L_0}; \vr y_{L_0}) \leq  h_0
\end{equation}
Combining Eq.(\ref{eq:rate_analysis2b}) and Eq.(\ref{eq:rate_analysis_delta}) we have:
\begin{equation}\label{eq:rate_analysis_delta2}
\Delta \leq \left(\frac{3 h_0}{K} + \frac{2}{n} \right) |\mathcal{X}| \cdot \log(n+1) - \frac{1}{n} \log \left( P_A \right)
\end{equation}
Combining Eq.(\ref{eq:rate_analysis2}),(\ref{eq:rate_analysis3}),(\ref{eq:rate_analysis4}) with Eq.(\ref{eq:convexity_in_rate_analysis}) and substituting $\Delta_m \leq \Delta_{\mu}$ yields:
\begin{multline}\label{eq:rate_analysis5}
 \hat I (\vr x; \vr y)  <
\Delta + \sum_{b=1}^{B+1}  \frac{m_b}{n} \left( \frac{K+ \Delta_{m_b}}{m_b}  \right) +
 \frac{B}{n} h_0
\leq \\ \leq
\Delta + \sum_{b=1}^{B+1}  \frac{1}{n} \left( K + \Delta_{\mu} \right) +
 \frac{B}{n} h_0
=\\=
\Delta + \frac{B+1}{n} \left( K + \Delta_{\mu}  \right) +
 \frac{B}{n} h_0
\end{multline}

From Eq.(\ref{eq:rate_analysis5}) $B$ and consequently $R_{\mathrm{act}}$ can be lower bounded:
\begin{multline}\label{eq:rate_analysis6}
R_{\mathrm{act}} = \frac{B}{n} \cdot K
 > \frac{\hat I (\vr x; \vr y) - \Delta - \frac{1}{n} \left( K + \Delta_{\mu} \right)}{K + \Delta_{\mu} + h_0} \cdot K
=\\= \frac{\hat I (\vr x; \vr y) - \Delta - \frac{K}{n} \left( 1 + \frac{\Delta_{\mu}}{K} \right)}{1 + \frac{\Delta_{\mu} + h_0}{K}}
\end{multline}

Now if we increase $K$ with $n$ such that $O(\log(n))<O(K)<O(n)$ (for example by choosing $K=n^\alpha$, $0<\alpha<1$), then $\frac{K}{n} \to 0$ as $n \to \infty$, since $\Delta_{\mu} = O(\log(n))$ we have $\frac{\Delta_{\mu}}{K} \to 0$ and from Eq.(\ref{eq:rate_analysis_delta2}) we have $\Delta \to 0$ thus for any $\epsilon$ we have $n$ large enough so that:
\begin{multline}
R_{\mathrm{act}} > \frac{\hat I (\vr x; \vr y) - \epsilon}{1 + \epsilon}
> \left( \hat I (\vr x; \vr y) - \epsilon \right) (1 - \epsilon)
>\\>
\hat I (\vr x; \vr y) - (1+h_0) \epsilon
\equiv \Remp
\end{multline}
Outside the set $J$, where the last inequality is due to the fact $\hat I$ is bounded. Hence we proved our claim that the rate exceeds a rate function which converges uniformly to the empirical mutual information and the proof of Theorem \ref{theorem:discrete_adaptive} is complete.
\myendofproof

\subsubsection{Rate analysis for the continuous case}\label{sec:rate_analysis_continuous}
The continuous case is more difficult from several reasons. One is that the error probability exponent has a missing degree of freedom ($\approx \exp((n-1)t)$). This results in a rate loss (through $s$ in the definition of $\mu^*_m$), which is larger for small blocks, and can be bounded only when assuming the number of blocks does not grow linearly with $n$. Since the effective mutual information $\Remp(\vr x, \vr y)$ is unbounded we cannot simply bound the loss of mutual information over the unused symbols. Specifically for a single symbol, $\hat\rho=1$ and $\Remp=\infty$. Therefore we use the convexity of the correlation factor and the fact it is bounded by 1. As a result, the loss introduced in order to attain convexity (over the rateless blocks) is in the correlation factor rather than the empirical mutual information. A loss in the correlation factor induces unbounded loss in the rate function for $\rho \approx 1$, leading to a maximum rate. In order to cope with these difficulties we use a threshold $T$ on the number of symbols in a block ($T$ is chosen to grow slower than $n$), and treat large and small blocks differently: the large blocks are analyzed through their correlation factor and for the small blocks the correlation factor is upper bounded by 1 and only the number of blocks is accounted for.


We denote $\hat\rho_b \equiv \hat\rho(\vr x_{U_b}, \vr y_{U_b})$ and $\hat\rho \equiv \hat\rho(\vr x, \vr y)$ the correlation factor measured on a rateless block and on the entire transmission block, respectively. We denote by $B_S = \{b : m_b \leq T\}$ and $B_L = \{b: m_b > T\}$ the indices of the small and the large blocks respectively (the last unfinished block included). The total number of symbols in the large blocks is denoted $m_L \equiv \sum_{b \in B_L} m_b$. The number of large blocks is bounded by $|B_L| < \frac{n}{T}$.

The decoding threshold is written as
\begin{equation}
\mu^*_m = \frac{K}{m-1} + \frac{1}{m-1}\log \left( \frac{n}{P_e} \right) + \frac{\log(2)}{m-1} = \frac{K + \Delta_{\mu}}{m-1}
\end{equation}
where we denoted $\Delta_{\mu} \equiv \log \left( \frac{2n}{P_e} \right)$. We consider the partitioning of the index set $\{1,\ldots,n\}$ into at most $p=\frac{n}{T}$ sets: the first $\frac{n}{T}-1$ (or less) sets are the large blocks except their last symbol $\bigcup_{b \in B_L} U_b$ (each with at least $T+1$ symbols by definition), and the last set denoted $L_1$ includes the rest of the symbols (last symbols of these blocks and all symbols of small blocks), and has $|L_1| = n - m_L$. Since this partitioning has a bounded number of sets, by applying Lemma \ref{lemma:likely_convexity_of_rho} and Eq.(\ref{eq:likely_convexity_with_dependency}) with $p = \frac{n}{T}$ we have that Eq.\ref{eq:convexity_in_cont_rate_analysis} below is satisfied when $\vr x$ is outside a set $J$ with
probability at most:
\begin{multline}\label{eq:convexity_prob_in_cont_rate_analysis}
\Pr(J) \leq n^{2p} \cdot 2^p e^{ - n \Delta^2/8}
= \left( \sqrt{2} n \right)^{2\frac{n}{T}} e^{ - n \Delta^2/8}
=\\=
\exp \left[  - n \left( \log(e) \Delta^2/8 - \frac{2}{T} \log \left( \sqrt{2} n \right)  \right) \right]
\end{multline}
For any $0 < \Delta \leq \frac{1}{7}$. This bound tends to 0 if $T > O(\log(n))$ (since $\log(e) \Delta^2/8 - \frac{2}{T} \log \left( \sqrt{2} n \right) \to \log(e) \Delta^2/8 > 0$) therefore for any such $\Delta$ there is $n$ large enough such that this probability falls
below the required $P_A$. The convexity condition is:
\begin{multline}\label{eq:convexity_in_cont_rate_analysis}
\hat \rho^2 - \Delta \leq  \sum_{b \in B_L} \frac{m_b}{n} \hat \rho_b^2 +
 \frac{|L_1|}{n} \hat \rho(\vr x_{L_1}; \vr y_{L_1})^2
\leq \\ \leq
 \sum_{b \in B_L} \frac{m_b}{n} \hat \rho_b^2 + \frac{n - m_L}{n}
 \end{multline}
where $\Delta$ can be made arbitrarily close to 0. We define a factor $\eta_1<1$ and apply the function $\smallminushalf \log(1-\eta_1 t)$ to both sides of the above equation. Since the function is monotonically increasing and convex  $\cup$ over $t \in [0,1)$ (stemming from  concavity $\cap$ of $\log(t)$), we have:
\begin{multline}\label{eq:cont_rate_analysis3}
r_0 \equiv \smallminushalf \log(1 - \eta_1 \cdot (\hat \rho^2 - \Delta))
\leq \\ \stackrel{\text{(\ref{eq:convexity_in_cont_rate_analysis})}}{\leq} \smallminushalf \log \left[ 1 - \eta_1 \left( \sum_{b \in B_L} \frac{m_b}{n} \hat \rho_b^2 + \frac{n - m_L}{n} \cdot 1\right) \right]
 \leq \\ \leq
 \sum_{b \in B_L} \frac{m_b}{n} \smallminushalf \log \left( 1 - \eta_1 \hat \rho_b^2 \right)
+\\+ \frac{n - m_L}{n} \smallminushalf \log \left( 1 - \eta_1 \cdot 1 \right)
 \end{multline}

We start by bounding the terms related to the large blocks. At the last symbol before decoding in each block (or symbol $n$ for the unfinished block) none of the codewords, including the correct one crosses the threshold $\mu_m^*$, therefore we have for $b=1,\ldots,B+1$:
\begin{equation}\label{eq:cont_rate_analysis2}
\mu^*_{m_b} = \frac{K + \Delta_{\mu}}{m_b-1} > \Remp(\vr x_{U_b}, \vr y_{U_b}) = -\half \log (1 - \hat\rho_b^2)
\end{equation}
and since $m_b \geq T+1$:
\begin{multline}\label{eq:cont_rate_analysis4}
\frac{m_b}{n} \smallminushalf \log \left( 1 - \eta_1 \hat \rho_b^2 \right) \leq
\frac{m_b}{n} \smallminushalf \log \left( 1 - \hat \rho_b^2 \right)
< \\ \stackrel{\text{(\ref{eq:cont_rate_analysis2})}}{<}
\frac{m_b}{n} \cdot \frac{K + \Delta_{\mu}}{m_b-1} = \left(1 + \frac{1}{m_b-1} \right) \frac{K + \Delta_{\mu}}{n}
\leq \\ \leq
\left(1 + \frac{1}{T}\right) \frac{K + \Delta_{\mu}}{n}
 \end{multline}
For the small blocks we use $n \leq \sum_{b \in B_L} (m_b + 1) + \sum_{b \in B_S} (m_b + 1) \leq m_L + |B_L| + (T+1) |B_S|$ (where the inequality is since the unterminated block has length $m_b$) to bound
$ n - m_L \leq  |B_L| + (T+1) |B_S|$.

Combining Eq.(\ref{eq:cont_rate_analysis3}) with these bounds we have:
\begin{multline}\label{eq:cont_rate_analysis5}
r_0 \leq
 |B_L|  \left(1 + \frac{1}{T}\right) \frac{K + \Delta_{\mu}}{n}
+\\+
\frac{|B_L| + (T+1) |B_S|}{n}  \left[ - \half \log \left( 1 - \eta_1 \right) \right]
\end{multline}
The last equation is a lower bound on a linear combination of $|B_L|$ and $|B_S|$. Since the total information sent depends on $|B_L| + |B_S|$ we equalize the coefficients multiplying $|B_L|$ and $|B_S|$ by determining $\eta_1$ so that:
\begin{equation}\label{eq:cont_rate_analysis6}
 - \half \log \left( 1 - \eta_1 \right) =  \left(1 + \frac{1}{T}\right) \frac{K + \Delta_{\mu}}{T}
\end{equation}
This is always possible since the RHS is positive and the LHS maps $\eta_1 \in (0,1)$ to $(0, \infty)$. Then
\begin{multline}\label{eq:cont_rate_analysis7}
r_0 \leq  \left(|B_L| + \frac{|B_L| + (T+1) |B_S|}{T} \right) \left(1 + \frac{1}{T}\right) \frac{K + \Delta_{\mu}}{n}
=\\=
\left(|B_L| + |B_S| \right)  \frac{K + \Delta_{\mu}}{n} = \left(B+1 \right)  \frac{K + \Delta_{\mu}}{n}
\end{multline}
Extracting a lower bound on $B$ from Eq.(\ref{eq:cont_rate_analysis7}) yields a bound on the empirical rate:
\begin{multline}\label{eq:cont_rate_analysis8}
R_{\mathrm{act}} = \frac{K}{n} \cdot B
\geq \\ \geq \frac{K}{n} \cdot \left( \frac{r_0 \cdot n}{K + \Delta_{\mu}} -1 \right) =
\frac{r_0}{ 1 + K^{-1}\Delta_{\mu}} -\frac{K}{n}
=\\=
\frac{\smallminushalf \log(1 - \eta_1 (\hat \rho^2 - \Delta))}{ (1 + K^{-1}\Delta_{\mu})} -\frac{K}{n}
\equiv \RLBONE
\end{multline}
Equation (\ref{eq:cont_rate_analysis8}) may be optimized with respect to $T$ to obtain a tighter bound, but this is not necessary to prove the theorem. Recall that $\Delta_{\mu} = O(\log(n))$. By choosing $O(\log(n)) < K < O(n)$ the factor $(1 + K^{-1}\Delta_{\mu})$ in Eq.(\ref{eq:cont_rate_analysis8}) can be made arbitrarily close to 1 and $\frac{K}{n}$ can be made arbitrarily close to 0. As we saw above choosing $O(\log(n)) < T < O(n)$ enables us to
have $P_A \to 0$ with $\Delta$ arbitrarily close to 0, and finally if $K > O(T)$ then the RHS of Eq.(\ref{eq:cont_rate_analysis6}) tends to $\infty$ and therefore we can choose $\eta_1$ arbitrarily close to 1. Summarizing the above, by selecting $O(\log(n)) < O(T) < O(K) < O(n)$ we can write the rate as
\begin{equation}\label{eq:cont_rate_analysis9}
R_{\mathrm{act}} \geq \RLBONE = \smallminushalf \log(1 - \eta_1 \cdot (\hat \rho^2 - \Delta)) \cdot \eta_2 - \epsilon_1
\end{equation}
With $\eta_1, \eta_2 \arrowexpl{n \to \infty} 1^{-}$ and $\epsilon_1,\Delta \arrowexpl{n \to \infty} 0^+$.
$\RLBONE$ tends to the target rate $ R_2(\hat\rho) \equiv \half \log \left( \frac{1}{1 - \hat\rho^2} \right)$ for each point $\hat\rho \in [0,1)$ (but not uniformly), and it remains to show that for any $\bar{R}, \epsilon$ there is $n$ large enough such that $\RLBONE \geq \RLBTWO \equiv \min(R_2(\hat\rho)-\epsilon, \bar{R})$.

The functions $R_2(\rho)$ and $\RLBONE(\rho)$ are monotonically increasing (for fixed $\eta_1, \eta_2$ and $\epsilon_1$) and it is easy to verify by differentiation that the difference $R_2(\rho) - \RLBONE(\rho)$ is also monotonically increasing. Given $\bar{R}, \epsilon$, choose $\rho_0$ such that $R_2(\rho_0) = \bar{R} + \epsilon$. Since $\RLBONE(\rho_0) \arrowexpl{n \to \infty} R_2(\rho_0)$, for $n$ large enough we have $R_2(\rho_0) - \RLBONE(\rho_0) \leq \epsilon$, and therefore $\RLBONE(\rho_0) \geq R_2(\rho_0) - \epsilon = \bar{R}$. For this $n$, for any $\rho \leq \rho_0$ from the monotonicity of the difference we have that $R_2(\rho) - \RLBONE(\rho) \leq \epsilon$, and for any $\rho \geq \rho_0$ we have from the monotonicity of $\RLBONE(\rho)$ that $\RLBONE(\rho) \geq \bar{R}$, therefore $\RLBONE \geq \RLBTWO$, which completes the proof of Theorem \ref{theorem:continuous_adaptive}.
\myendofproof

\section{Examples}\label{sec:examples}
In this section we give some examples to illustrate the model developed in this paper. In this section we use a slightly less formal notation.

\subsection{Constant outputs and other illustrative cases}\label{sec:paradox}
The statement that a rate which is determined by the input and output sequences can be attained without assuming any dependence between them may seem paradoxical at first. Some insight can be gained by looking at the specific case where the output sequence is fixed and does not depend on the input. In this case, obviously, no information can be transferred. Since the encoder uses random sequences, the result of fixing the output is that the probability to have an empirical mutual information larger than $\epsilon>0$ tends to $0$, therefore most of the time the rate will be $0$. Infrequently, however, the input sequence accidentally has empirical mutual information larger than $\epsilon>0$ with the output sequence. In this case the decoder will set a positive rate, but very likely fail to decode. These cases occur in vanishing probability and constitute part of the error probability. So in this case we will transmit rate $R=0$ with probability of at least $1-P_e$ and $R>0$ with probability at most $P_e$. Conversely, if the channel appears to be good according to the input and output sequences (suppose for example $y_k=x_k$), the decoder does not know if it is facing a good channel or just a coincidence, however it takes a small risk by assuming it is indeed a good channel and attempting to decode, since the chances of high mutual information appearing accidentally are small (and uniformly bounded for all output sequences).


Another point that appears paradoxical at first sight is that the decoder is able to determine a rate $R \geq \Remp$ without knowing $\vr x$ for any $\vr x \not\in J$. First observe that although it is an output of the decoder,
the rate $R$ is not controlled by the encoder and therefore cannot convey information. Since the decoder knows the codebook, and given the codebook
the sequence $\vr x$ is limited to a number of possibilities (determined by the possible messages and block locations), it is easy to find an $R(\vr y) \geq \Remp(\vr x, \vr y)$ by maximizing $\Remp$ over all possible sequences $\vr x$. Vaguely speaking, the decoding process is indeed a maximization of $\Remp$ over multiple $\vr x$ sequences and by Lemmas \ref{lemma:pairwise_discrete}, \ref{lemma:pairwise_continuous} such a decoding process guarantees small probability of
error.

\subsection{Applying the continuous alphabet scheme to other input alphabets}\label{sec:example_adaptation_func}
The scheme used for the continuous case can be adapted to peak limited or even discrete input, by using an adaptation function, i.e. the channel input will be $x_k' = f(x_k)$. In this case the modified codebook $C' = f(C)$ will be generated by passing the Gaussian codebook through the adaptation function, but for analysis purposes the adaptation function $f(\cdot)$ may be considered part of the channel and the correlation factor is calculated with respect to $\vr x$ which is used to generate the codebook. In order to write the rate guaranteed by this approach as a function of $\vr x'$ rather than $\vr x$, the law of large numbers has to be utilized (in general) with respect to the distribution $\Pr(x_k \vert x_k')$.

\subsection{Non linear channels}\label{sec:example_non_linear}
\def\Neff{N_{\mathrm{eff}}}
\def\Peff{P_{\mathrm{eff}}}
In analyzing probabilistic channels, the correlation model determines the rate $\half \log \left( \frac{1}{1-\rho^2} \right)$ is always achievable using Gaussian code (no randomization is needed if the channel is probabilistic as can be shown by the standard argument about the existence of a good code). This is actually a result of Lemma \ref{lemma:gaussian_mi_bound}.

This expression is useful for analyzing channels in which the noise is not additive or non linearities exist.  As an example, transmitter noise is usually modeled as an additive noise. However large part of this noise is due to distortion (e.g. in the power amplifier), and therefore depends on the transmitted signal and is inversely correlated to it. Consider the non linear channel $Y = f(X) + V$ with $V \sim \Normal(0,N)$. In this case if we define the effective SNR as $\textit{SNR}=\frac{\rho^2}{1-\rho^2}$ then rate $R=\half \log \left( 1 + \textit{SNR} \right)$ is achievable. The correlation factor is:
\begin{equation}
\rho^2 = \frac{E(XY)^2}{E(X^2) E(Y^2)} = \frac{E(X f(X))^2}{E(X^2) (E(f(X)^2) + N)}
\end{equation}
Therefore the effective \textit{SNR} is:
\begin{multline}\label{eq:eff_noise}
\textit{SNR}=\frac{\rho^2}{1-\rho^2}
=\\= \frac{E(X f(X))^2}{E(X^2) (E(f(X)^2) + N) - E(X f(X))^2} =
 \frac{\Peff}{ N + \Neff}
\end{multline}
where we defined the effective gain $\gamma$, the effective power $\Peff$ and the effective noise $\Neff$ as:
\begin{eqnarray}
\gamma &\equiv& \frac{E(X f(X))}{E(X^2)} \\
\Peff &\equiv& \frac{(E[(X f(X)])^2}{E(X^2)} = E\left[(\gamma X)^2\right] \\
\Neff &\equiv& E(f(X)^2) - \frac{(E[X f(X)])^2}{E(X^2)} \nonumber \\
&=& E \left[(f(X) - \gamma X)^2 \right]
\end{eqnarray}
This yields a simple characterization of the degradation caused by the non linearity, which is independent of the noise power and is tight if the non linearity is small. This model enables to characterize the transmitter distortions by the two parameters $\Peff, \Neff$, a characterization which is more convenient and practical to calculate than the channel capacity, and on the other hand guarantees that transmitter noise evaluated this way never degrades the channel capacity in more than determined by Eq.(\ref{eq:eff_noise}).

Another interesting application of this bound is in treating receiver estimation errors, since it is simpler to calculate the loss in the correlation factor induced due to the imperfect knowledge of the channel parameters than the loss in capacity. For example, the bound in \cite{Hassibi} for the loss due to channel estimation from training, when specialized to single input single output (SISO) channels, may be computed using the correlation factor bound.

\subsection{Employing continuous channel scheme over a BSC}\label{sec:example_continuous_over_BSC}
When operated over a channel different than the Gaussian additive noise channel, the rates achieved with the scheme we described in the continuous case are suboptimal compared to the channel capacity. The loss depends on the channel in question. As an example, suppose the communication system is used over a BSC with error probability $\epsilon$, i.e. the continuous input value $X$ is translated to a binary value by $\sign(X)$, and the output is $Y = \sign(X) \cdot (-1)^{Ber(\epsilon)}$. The capacity of this channel is $C = 1_{\mathrm{bit}}-h_b({\epsilon})$ and we are interested to calculate the rate which would be achieved by our scheme (which does not know the channel) for this channel behavior. For this channel with Gaussian $\Normal(0,P)$ input we have (through a simple calculation):
\begin{equation}
E(XY) = (1-2\epsilon) \sqrt{\frac{2P}{\pi}}
\end{equation}

\begin{figure}
\center
  \includegraphics[width=0.5\textwidth]{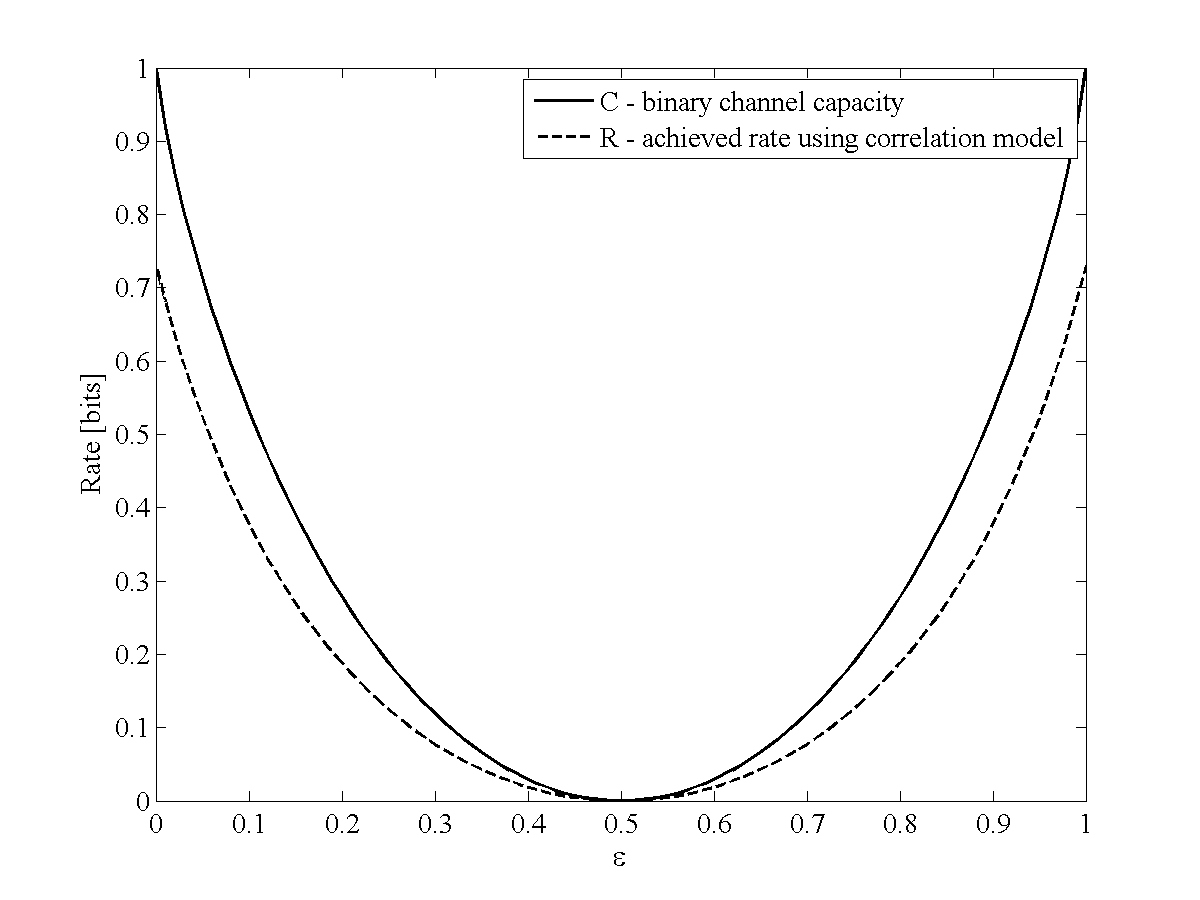}
  \caption{Comparison of C,R for the BSC}\label{fig:rate_in_bsc}
\end{figure}

Hence
\begin{equation}
\rho^2 = \frac{E(XY)^2}{P \cdot E(1^2)} = \frac{2}{\pi} (1-2\epsilon)^2
\end{equation}
And
\begin{equation}
R = \half \log \left( \frac{1}{1-\frac{2}{\pi} (1-2\epsilon)^2} \right)
\end{equation}
The comparison between $C$ and $R$ is presented in fig.(\ref{fig:rate_in_bsc}). It can be shown that $R \geq \frac{2}{\pi} C$, thus the maximum loss is 36\%.

\subsection{Channels that fail the zero order and the correlation model}\label{sec:example_failure_models}
Although we did not assume anything about the channel, and specifically we did not assume the channel is memoryless, the fact we used the zero-order empirical distribution means the results are less tight for channels with memory. Specifically if delay is introduced then the scheme would fail completely. For example, for the channel $y_k = x_k + \half x_{k-1} + v_k$ we would obtain positive rates and the intersymbol interference (ISI) $\half x_{k-1}$ would be treated (suboptimally) as noise, but for the error free channel $y_k = x_{k-1}$ the achieved rate would be 0 (with high probability). Similarly we can find a memoryless channel with infinite capacity but for which the correlation model we used for the continuous alphabet scheme fails: if $y_k = x_k^2$ then $\rho=0$. Another example of practical importance is the fading channel (with memory) $y_n=h_n x_n+v_n$, where $h_n$ is slowly fading with mean $0$. All these examples result from the simplicity of the models used, and can be solved by schemes employing higher order empirical distributions (over blocks, or by using Markov models), and by employing tighter approximations of the empirical statistics (e.g. by higher order statistics) in the continuous case.

\subsection{Using individual channel model to analyze adversarial individual sequence}\label{sec:example_adversary}
As we noted in the overview, the results obtained for the individual channel model constitute a convenient starting point for analyzing channel models which have a full or partial probabilistic behavior. It is clear that results regarding achievable rates in fully probabilistic, compound, arbitrarily varying and individual noise sequence models can be obtained by applying the weak law of large numbers to the theorems discussed here (limited, in general, to the randomized encoders regime).

E.g. for a compound channel model $W_{\theta}(y|x)$ with an unknown parameter $\theta$ since $\hat P(\vr x; \vr y) \arrowexpl{n \to \infty} P_{\theta}(x,y) = W_{\theta}(y|x) Q(x) $ in probability for every $\theta$ and since $I(\cdot ; \cdot)$ is continuous $\hat I(\vr x; \vr y) \arrowexpl{n \to \infty} I_{\theta}(X;Y)$. Hence from Theorem \ref{theorem:discrete_nonadaptive} rate $\min_{\theta} I_{\theta}(X;Y)$ can be obtained without feedback, and from Theorem \ref{theorem:discrete_adaptive} rate $I_{\theta}(X;Y)$ can be obtained with feedback. These results are not new (see \cite{Blackwell_compound}\cite{Lapidoth_compound} for the first and the second is obtained as a special case of the results in \cite{Eswaran} and \cite{Ofer_EMP} since the individual noise sequence model can be degenerated into a compound model) and are given only to show the ease of using the individual model once established.

To show the strength of the model we analyze a problem considered also in \cite{Ofer_EMP} of an individual sequence which is determined by an adversary and allowed to depend in a fixed or randomized way on the past channel inputs and outputs. For simplicity we start with the binary channel $y_k = x_k \oplus e_k$ where $e_k$ is allowed to depend on $\vr x_1^{k-1}$ and $\vr y_1^{k-1}$ (possibly in a random fashion), and the target is to show the empirical capacity is still achievable in this scenario. Note that here $E_k$ is a random variable but not assumed to be i.i.d. We denote the relative number of errors by $\hat \epsilon \equiv \frac{1}{n} \sum_{k=1}^n e_k$. We would like to show the communication scheme achieves a rate close to $1_{\mathrm{bit}} - h_b(\hat\epsilon)$ in high probability, regardless of the adversary's policy. Note that both the achieved rate and the target $1_{\mathrm{bit}} - h_b(\hat\epsilon)$ are random variables and the claim is that they are close in high probability (i.e. that the difference converges in probability to $0$ when $n \to \infty$)


Applying the scheme achieving Theorem \ref{theorem:discrete_adaptive} with $Q = Ber(\half)$ we can asymptotically approach (or exceed) the rate:
\begin{multline}
\hat I (\vr x; \vr y) = \hat H (\vr y) - \hat H (\vr y | \vr x) = \hat H (\vr y) - \hat H (\vr e | \vr x)
\geq \\ \geq
\hat H (\vr y) - \hat H (\vr e) = \hat H (\vr y) - h_b(\hat \epsilon)
\end{multline}
Note that unlike in the probabilistic BSC where we have $I(X;Y)=H(Y)-H(E)$, here the empirical distribution of $\vr e$ is not necessarily independent of $\vr x$, therefore the entropies are only related by the inequality $\hat H (\vr e | \vr x) \leq \hat H (\vr e)$ (conditioning reduces entropy). In order to show a rate of $1_{\mathrm{bit}} - h_b(\hat\epsilon)$ is achieved, we only need to show $\hat H (\vr y) \arrowexpl{n \to \infty, prob.} 1_{\mathrm{bit}}$. Since $X_k$ is independent of $X_1^{k-1}, Y_1^{k-1}$ and therefore also of $E_k$ we have:
\begin{multline}
\Pr(Y_k=0 | Y_1^{k-1}) = \sum_{e_k} \Pr(Y_k=0 | Y_1^{k-1}, e_k) \Pr(e_k)
=\\= \sum_{e_k} \Pr(X_k = e_k | Y_1^{k-1}, e_k) \Pr(e_k)
=\\= \sum_{e_k} \Pr(X_k = e_k) \Pr(e_k)
= \sum_{e_k} \half \Pr(e_k) = \half
\end{multline}
Therefore $Y_1^n$ is distributed i.i.d. $Ber(\half)$ and from the law of large numbers and the continuity of $H(\cdot)$ we have the desired result. This result is a special case of the results in \cite{Ofer_EMP}.

We can extend the example above to general discrete channels and perform a consolidation of the adversarial sequence model considered in \cite{Ofer_EMP} (for modulu additive channels) with the general discrete channel with fixed sequence considered in \cite{Eswaran}. We address the channel $W_s(y|x)$ with state sequence $s_k$ potentially determined by an adversary knowing all past inputs and outputs. We would like to show that the rate $I(Q, \sum_s W_s(y|x) \hat P_{\vr s}(s) )$ (the mutual information of the state-averaged channel) can be asymptotically attained in the sense defined above.

This result is a superset of the results of \cite{Eswaran} and \cite{Ofer_EMP}. It overlaps with \cite{Eswaran} in the case $\vr s$ is a fixed sequence and with \cite{Ofer_EMP} for the case of modulu-additive channel (or when the target rate is based on the modulu additive model).

Since Theorem \ref{theorem:discrete_adaptive} shows the rate $\hat I (\vr x; \vr y) \equiv I(\hat P(\vr x), \hat P(\vr y | \vr x))$ can be approached or exceeded asymptotically, it remains to show that the empirical distribution $\hat P(\vr x, \vr y)$ is asymptotically close to the state-averaged distribution $P_{avg}(x,y) \equiv \sum_s W_s(y|x) \hat P_{\vr s}(s) Q(x) = \frac{1}{n} \sum_k W_{S_k}(y|x) Q(x) $, and the result will follow from continuity of the mutual information. Note that the later value is a random variable (function) depending on the behavior of the adversary. Here we do not use the law of large numbers because of the interdependencies between the signals $\vr x, \vr y$ and $\vr s$.

Our purpose is to prove that the difference $\Delta(t,r)$ defined below converges in probability to 0 for every $t,r$:
\begin{multline}
\Delta(t,r) \equiv \hat P_{(\vr x, \vr y)}(t,r) - P_{avg}(t,r)
=\\= \frac{1}{n} \sum_k \Ind(X_k = t, Y_k = r) - \frac{1}{n} \sum_k W_{S_k}(r|t) Q(t)
\equiv \\ \equiv
\frac{1}{n} \sum_k \varphi_k(t,r)
\end{multline}
where $\varphi_k(t,r) \equiv \Ind(X_k = t, Y_k = r) - W_{S_k}(r|t) Q(t)$. For brevity of notation we omit the argument $(t,r)$ from $\varphi_k(t,r)$ since from this point on it takes a fixed value.
Then
\begin{multline}\label{eq:example_adversary1}
E (\Ind(X_k = t, Y_k = r) \vert X^{k-1},Y^{k-1},S^{k})
=\\= \Pr(X_k = t, Y_k = r \vert X^{k-1},Y^{k-1},S^{k})
=\\= \Pr(X_k = t \vert X^{k-1},Y^{k-1},S^{k})
  \cdot \\ \cdot \Pr(Y_k = r \vert X_k = t, X^{k-1},Y^{k-1},S^{k})
=\\ \stackrel{(a)}{=}
\Pr(X_k = t) \cdot \Pr(Y_k = r \vert X_k = t, S_k)
\stackrel{(b)}{=}  Q(t) W_{S_k}(r|t)
\end{multline}
where (a) is due to the independent drawing of $X_k$ (when not conditioned on the codebook), the fact $S^k$ is independent of $X_k$, and the memoryless channel (defining the Markov chain $(X^{k-1}, Y^{k-1}, S^{k-1}) \leftrightarrow (X_k,S_k) \leftrightarrow Y_k$), and (b) is due to the i.i.d drawing of $X_k$ from $Q$ and the definition of $W$. From Eq.(\ref{eq:example_adversary1}) we have that:
\begin{equation}
E(\varphi_k \vert X^{k-1},Y^{k-1},S^{k}) = 0
\end{equation}

By the smoothing theorem we also have that $\varphi_k$ has zero mean $E(\varphi_k)=0$. We now show that $\varphi_k$ are uncorrelated. Consider two different indices $j < k$ (without loss of generality) then
\begin{multline}
E(\varphi_k\cdot \varphi_j) = E \left[ E(\varphi_k \cdot \varphi_j \vert X^{k-1},Y^{k-1},S^{k}) \right]
=\\=
E \left[ \varphi_j \cdot E(\varphi_k \vert X^{k-1},Y^{k-1},S^{k}) \right] = 0
\end{multline}
where we used the smoothing theorem and the fact $\varphi_j$ is completely determined by $X_j,Y_j,S_j$ which are given. In addition since by definition $-1 \leq \varphi_k \leq 1$, $E(\varphi_k^2) \leq 1$. Therefore
\begin{equation}
E (\Delta^2) = \frac{1}{n^2} \sum_{j,k=1}^n E( \varphi_k \cdot \varphi_j) \leq \frac{1}{n^2} \sum_{j,k=1}^n \delta_{jk} = \frac{1}{n}
\end{equation}
and by Chebychev inequality for any $\epsilon>0$:
\begin{equation}
\Pr(|\Delta(t,r)| > \epsilon) \leq \frac{E (\Delta^2)}{\epsilon^2} \leq \frac{1}{n \epsilon^2} \arrowexpl{n \to \infty} 0
\end{equation}
which proves the claim. \myendofproof

This result is new, to our knowledge, however the main point here is the relative simplicity in which it is attained when relying on the empirical channel model (note that most of the proof did not require any information-theoretic argument).

\section{Comments and further study}\label{sec:comments}
\subsection{Limitations of the model}
The scheme presented here is suboptimal when operated over channels with memory or, in the continuous case over non AWGN channels, and in section \ref{sec:example_failure_models} we discussed several cases where the communication fails completely. Obviously the solution is to extend the time order of the model. A simple extension is by using the super-alphabets $\mathcal{X}^p$ and $\mathcal{Y}^p$ and treating a block of channel uses as one symbol. A more delicate extension is by considering a Markov model (the $p$-th order empirical conditional probability $\hat P (x_k,y_k | \vr x_{k-p}^{k-1}, \vr y_{k-p}^{k-1})$).

For the continuous channel we focused on a specific class of continuous channels where the alphabet is the real numbers (we have not considered vectors as in MIMO channels), and we did not achieve the full mutual information. A possible extension is to find measures of empirical mutual information for the continuous channels which are also attainable and approach the probabilistic mutual information for probabilistic channels. The current paper exhibits a considerable similarity between the continuous case and the discrete case which is not fully explored here, and a unifying theory which will include the two as particular cases is wanting.

We conjecture that the following definition of empirical mutual information may achieve these goals: given a family of joint distributions (not necessarily i.i.d) $\{ P_\theta(\vr x, \vr y), \theta \in \Theta\}$ define the entropy with respect to the family $\Theta$ as the entropy of the closest member of the family (in maximum likelihood sense): $\hat H_{\Theta}(\vr x) = \min_{\theta \in \Theta} - \frac{1}{n} \log P_{\theta}(\vr x)$ and likewise $\hat H_{\Theta}(\vr x|\vr y) = \min_{\theta \in \Theta} - \frac{1}{n} \log P_{\theta}(\vr x | \vr y)$, and define the relative mutual information as $\hat I_{\Theta}(\vr x;\vr y) = \hat H_{\Theta}(\vr x) - \hat H_{\Theta}(\vr x|\vr y)$. This definition corresponds to our target rates for the discrete case (with $\Theta$ as the family of all DMC-s) and continuous case (with $\Theta$ the family of all joint Gaussian zero-mean distributions $\Normal(0,\Lambda_{XY})$).

\subsection{Overhead and error exponent}
Another aspect is the overhead associated with extending the empirical distribution ("channel") family which is considered (both in considering time dependence and in increasing the accuracy with which the distribution is estimated or described). This overhead is related to the redundancy or regret associated with universal distributions (see \cite{Barron_MDL}). Although we haven't performed a detailed analysis of the overheads and considered only the asymptotically achievable rates, it is obvious from comparing Lemmas \ref{lemma:pairwise_discrete} and \ref{lemma:pairwise_continuous} that the tighter rates we obtained for the discrete channel come at the cost of additional overhead ($O(\log(n))$ compared to $O(1)$ in the continuous case) which is associated with the richness of the channel family (describing a conditional probability as opposed to a single correlation factor). Thus for example for a discrete channel with a large alphabet and a small block size $n$ we would sometimes be better off using the "continuous channel model" version of our scheme (gaining only from the correlation) rather than the scheme of the discrete case (gaining the empirical mutual information). The issue of overheads requires additional analysis in order to determine the bounds on the overheads and the tradeoff between richness of the channel family and the rate, for a finite $n$. As we noted in section \ref{sec:rate_analysis_continuous} the bounds we currently have for the rate-adaptive continuous case are especially loose and call for improvement.

Since rate can be traded off for error probability, a related question is the error exponent. Here, a good definition is still lacking for variable rate schemes, and the error exponents are not known for individual channels. The scheme we described does not endeavor to attain a good error exponent. Specifically, since the block of $n$ channel uses is broken into multiple smaller blocks, it is probably not an efficient scheme in terms of error rate. We note, however, that for rate adaptive schemes with feedback a good error exponent does not necessarily relate to the capability of sending a message with small probability of error, but rather to the capability to detect the errors. A similar situation occurs in the setting of random decision time considered by Burnashev \cite{Burnashev}. In the later, an uncertainty of the decoder with respect to the message is mitigated by sending an acknowledge / unacknowledge (ACK/NACK) message and possibly repeating the transmission with small penalty in the average rate (see a good description in \cite{Tchamkerten} sec IV.B). A similar approach can be used in our setting (fixed decoding time, variable rate), by sending an ACK/NACK over a fixed portion of the block and setting $R=0$ when the decoder is not certain of the received message. However we did not perform a detailed analysis. Note also that the analysis of the probability $P_A$ to transmit at a rate lower than the target rate function is entangled with the error analysis, since by such schemes it is possible to trade off rate for error, and reduce the error probability at the expense of increasing the probability to fall short of the target rate.

\subsection{Determining the behavior of the transmitted signal (prior)}
In this work we assumed a fixed prior (input probability distribution) and haven't dealt with the question of determining the prior, or more generally, how the encoder should adapt its behavior based on the feedback. Had the channel been a compound one, it stands to reason that a scheme using feedback may estimate the channel and adjust the input prior, and may asymptotically attain the channel capacity. However in the scope of individual channels (as well as individual sequence channels and AVC-s) it is not clear whether the approach of adjusting to the input distribution to the measured conditional distribution is of merit, if the empirical channel capacity can be attained for every sequence, and even the definition of achievability is unclear if the input distribution is allowed to vary.

Another related aspect is what we require from a communications system when considered under the individual channel framework. This question is relevant to all the requirements defined in the theorems (for example is the existence of the failure set $J$ necessary ?), however the most outstanding requirement is related to the prior.

Currently we constrained the input sequence to be a random i.i.d. sequence chosen from a fixed prior, which seems to be an overly narrow definition. The rationale behind this choice is that without any constraint on the input, the theorems we presented can be attained in a void way by transmitting only bad (e.g. fixed) sequences that guarantee zero empirical rate. Furthermore, without this constraint, attainability results for probabilistic models, and in general any attainable rates which are not conditioned on the input sequence could not be derived from our individual sequence theorems. A weaker requirement from the encoder is to be able to emit any possible sequence, however this requirement is not sufficient, since from the existence of such encoders we could not infer the existence of encoders achieving any positive rate over a specific channel. Consider for example the encoder satisfying the requirement by transmitting bad sequences in probability $1-\epsilon$ and good sequences in probability $\epsilon \to 0$. Theorems \ref{theorem:discrete_nonadaptive},\ref{theorem:continuous_nonadaptive},\ref{theorem:discrete_adaptive} and \ref{theorem:continuous_adaptive} are existence theorems, i.e. they guarantee the existence of at least one system satisfying the conditions. Had we removed the requirement for fixed input prior we saw these theorems would be attained by encoders that are unsatisfactory in other aspects. Once the theorem is satisfied by one encoder it cannot guarantee the existence of other (satisfactory) encoders, thus making it un-useful. Therefore the requirement for fixed prior is necessary in the current framework. Although in the scope of the theorems presented here, this requirement only strengthens the theorems (since it reveals additional properties of the encoder attaining the other conditions of the theorem), we are still bothered by the question what should be the minimal requirements from a communication system, and these hopefully will not include a constraint on the input distribution.

This issue relates to a fundamental difficulty which aries in communication over individual channels: unlike universal source coding in which the sequence is given a-priori, here the sequences are given a-posteriori, and the actions of the encoder affect the outcome in an unspecified way. Currently we broke the tie by placing a constraint on the encoder, but we seek a more general definition of the problem.

\subsection{Amount of randomization}
We have assumed so far there is no restriction on the amount of common randomness available and have not attempted to minimize the amount of randomization required (while maintaining the same rates). It is shown in \cite{Ofer_EMP} that less than $O(n)$ of randomization information is required in some cases and $O(n)$ is enough for others (see section V.5 therein), whereas we have used at least $O(M\cdot n) > O(n^2)$ random drawings to produce the codebook.

\subsection{Practical aspects}
The scheme described in this work is a theoretical one, but the concept appears to be extendable to practical coding systems. Below we focus on the continuous case and merely give the motivation (without proof). One may replace the correlation receiver (GLRT) by a receiver utilizing training symbols to learn the channel effective gain, and then apply maximum likelihood (or approximate, e.g. iterative) decoding. The randomization of the codebook may be replaced by using a fixed code with random interleaving, since with random interleaving only the empirical distribution of the (effective) noise sequence affects the error probability, and we may conjecture that the property that Gaussian noise distribution is the worst is approximately true for practical codes (such as turbo codes and LDPC). When using a random interleaver the training symbols as well as the part of the coded symbols can be interleaved together, and the decoding attempts (which occur every symbol in the theoretical scheme) occur only at the end of each interleaving block. The rateless code is replaced by an incremental redundancy scheme, i.e. by sending each time part of the symbols of the codeword, and repeating the codeword if all symbols were transmitted without successful decoding. The decision when to decode can be simply replaced by decoding and using a CRC check. Finally the common randomness (required only for the generation of the interleaver permutation) can be replaced by pseudo-randomness. Such a scheme may not be able to attain the promise of Theorem \ref{theorem:continuous_adaptive} for every individual sequence but may be able to adapt to every natural and man-made channel.

\subsection{Random decision time}
In our discussion we have described two communication scenarios: fixed rate without feedback and variable rate with feedback, and in both we assumed a fixed block size $n$. Another scenario is that of random decision time or rateless coding (as in \cite{Burnashev} \cite{Shulman}) in which the block size is not fixed but determined by the decoder. We did not include this scenario since the achievability result is less elegant in a way: the decoder indirectly affects the target rate (mutual information) through the block size. On the other hand this case may be of practical interest. Clearly the mutual information can be asymptotically attained for this communication scenario as well and its analysis is merely a simpler version of the rate analysis performed in section \ref{sec:rate_analysis}, since convexity is not required.

\subsection{Bounds}
In this paper we focused on achievable rates and did not show a converse. An almost obvious statement is that
any continuous rate function which depends only on the zero-order empirical statistics / correlation (respectively)
cannot exceed asymptotically the rate functions of Theorems \ref{theorem:discrete_adaptive}, \ref{theorem:continuous_adaptive}
respectively with vanishing error probability. To show the statement for the discrete case determine $\vr y$
using a memoryless channel $W(y|x)$. Then by the law of large numbers the empirical distribution
converges to the channel distribution and from the continuity of the rate function the empirical
rate converges to the rate function taken at the channel distribution. Since by Theorem
\ref{theorem:discrete_adaptive} the actual rate asymptotically meets or exceeds the rate function, and by the converse of the channel capacity theorem the actual rate cannot exceed (asymptotically) the mutual information, we have that
the rate function cannot exceed the mutual information ($\Remp \leq R_{act} \leq I(P,W)$), up to asymptotically vanishing factors. For the continuous case the analogue claim is shown by taking a Gaussian additive channel and replacing "distribution" by "correlation" and "empirical mutual information" by $-\half \log (1 - \hat\rho^2)$. The same applies also to rate functions obeying the conditions of Theorems \ref{theorem:discrete_nonadaptive}, \ref{theorem:continuous_nonadaptive}. More general bounds are yet to be studied.

\begin{table*}[t]
\caption{Comparison rate adaptive schemes in current paper and \cite{Eswaran}}\label{table:comparison_with_Eswaran}
\center
\begin{tabular}{|p{3cm}|p{4cm}|p{4cm}|p{4cm}|}
\hline
Item            & Eswaran  et al \cite{Eswaran}   &   Current Paper               & Comments \\ \hline
Channel model   & Individual sequence       & Individual channel            & \\ \hline
Mechanism for adaptivity & Repeated instanced of rateless coding & Repeated instanced of rateless coding & \\ \hline
Transmit format & Total time divided to rounds (=rateless blocks) which are divided to chunks & Total time divided to rateless blocks & Chunks in \cite{Eswaran} used as feedback instances and expurgated code has constant type over chunks \\ \hline
Feedback        & Ternary (Bad Noise/Decoded/Keep Going), once per chunk & Binary (Decoded/Not Decoded) per symbol & Easy to generalize to once every $1/\epsilon$ symbols (see \ref{sec:rate_analysis}) \\ \hline
Alphabet        & Discrete                  & Discrete or Real valued       & \\ \hline
Training        & Known symbols in random locations in each chunk & None    & \\ \hline
Randomness      & Full ($O(\exp(nR))$)      & Full ($O(\exp(nR))$)          & Might be reduced by selection from a smaller collection of codebooks (in both cases) \\ \hline
Codebook construction & Constant composition + expurgation + training insertion & Random i.i.d. & \\ \hline
Stopping condition & Threshold over mutual information of channel estimated from training & Threshold over empirical mutual information of best codeword & \\ \hline
Decoding & Maximum (empirical) mutual information & Maximum (empirical) mutual information & \\ \hline
Stopping location & End of Chunk & Any symbol & \\ \hline
\end{tabular}
\end{table*}

\subsection{Comparison of the rate adaptive scheme with the similar scheme in \cite{Eswaran}}
As noted the rate adaptive scheme we use is similar to the scheme of \cite{Eswaran} in its high level structure. Table \ref{table:comparison_with_Eswaran} compares some attributes of the schemes.

Another important factor is the overhead (i.e. the loss in number of bits communicated with a given error exponent, compared to the target rate), which we were unable to compare. We conjecture that the current scheme may have a lower overhead due to its simplicity which results in a smaller number of parameters and constraints on their order of magnitude (compared to the scheme of \cite{Eswaran} where relations between factors such as number of pilots and the minimum size of a chunk may require a large value of $n$).

\section{Conclusion}
We examined achievable transmission rates for channels with unspecified models, and focused on rates determined by a channel's a-posteriori empirical behavior, and specifically on rate functions which are determined by the zero-order empirical distribution. This communication approach does not require a-priori specification of the channel model. The main result is that for discrete channels the empirical mutual information between the input and output sequences is attainable for any output sequence using feedback and common randomness, and for continuous real valued channels an effective "Gaussian capacity" $-\half(1-\hat\rho^2)$ can be attained. This generalizes results obtained for individual noise sequences and is a useful model for analyzing compound, arbitrarily varying, and individual noise sequence channels.

\section*{Acknowledgment}
The authors would like to thank the reviewers of the ISIT 2009 conference paper on the subject for their helpful comments and references.



\appendix
\subsection{Proof of Lemma 1}\label{appendix:pairwise_discrete}
The proof is a rather standard calculation using the method of types. We use the notations of \cite{MethodOfTypes}. We divide the sequences according to their joint type $\mathcal{T}_{XY}$. The type $\mathcal{T}_{XY}$ is defined by the probability distribution $T_{XY} \in \mathcal{P}_n(\mathcal{XY})$. For notational purposes we define the dummy random variables $(\tilde{X},\tilde{Y}) \sim T_{XY}$ and $T_X$, $T_Y$, $T_{Y|X}$ as the marginal and conditional distributions resulting from $T_{XY}$. Following \cite{MethodOfTypes}, the conditional type is defined as $\mathcal{T}_{X|Y}(\vr y) \equiv \left\{\vr y : (\vr x, \vr y) \in \mathcal{T}_{XY}\right\}$. The empirical mutual information of sequences in the type $\mathcal{T}_{XY}$ is simply $I(\tilde{X};\tilde{Y}) = I(T_Y , T_{Y|X})$.
Define $T_t \equiv \{ T_{XY} \in \mathcal{P}_n(\mathcal{XY}): I(T_Y,T_{Y|X}) \geq t \}$. Since all sequences in the conditional type have the same (marginal) type, we can write:
\def\region1{\substack{T_{XY} \in \mathcal{P}_n(\mathcal{XY}):\\I(T_Y,T_{Y|X}) \geq t}}
\begin{multline}
Q^n \left( \hat{I}(\vr x; \vr y) \geq t \right) = \sum_{T_t} Q^n \left( \mathcal{T}_{X|Y}(\vr y) \right) = \\ \stackrel{(a)}{=}
\sum_{T_t} \lvert \mathcal{T}_{X|Y}(\vr y) \rvert \exp \left\{-n \left[H(T_X) + D(T_X || Q) \right]\right\}
\leq \\ \stackrel{(b)}{\leq}
\sum_{T_t} \exp \left\{ n H(\tilde{X} | \tilde{Y}) \right\} \exp \left\{-n \left[H(\tilde{X}) + D(T_X || Q) \right]\right\}
=\\=
\sum_{T_t} \exp \left\{-n \left[I(\tilde{X};\tilde{Y}) + D(T_X || Q) \right]\right\}
\leq \\ \leq
|\mathcal{P}_n(\mathcal{XY})| \cdot \exp \left\{-n \left( \min_{T_t} \left[I(T_Y,T_{X|Y}) + D(T_X || Q) \right] \right) \right\}
 \\ \stackrel{(c)}{\leq}
(n+1)^{|\mathcal{X}||\mathcal{Y}|} \cdot \exp \left(-n t \right)
=\\= \exp \left\{ -n \left(t - |\mathcal{X}||\mathcal{Y}|\frac{\log(n+1)}{n} \right) \right\}
\end{multline}
where (a) is due to \cite{MethodOfTypes} Eq.(II.1), (b) results from eq.(\ref{eq:size_of_condtype}) below which is an extension of (II.4) there to conditional types (and is a stronger version of Lemma II.3), based on the fact that in the conditional type $\mathcal{T}_{X|Y}(\vr y)$ the values of $\vr x$ over the $n_a = n T_Y(a)$ indices for which $y_i = a$ have empirical distribution $T_{X|Y}$ and therefore the number of such sequences is limited to $\exp \left( n_a H(\tilde{X} | \tilde{Y}=a) \right) $, hence:
\begin{multline}\label{eq:size_of_condtype}
\lvert \mathcal{T}_{X|Y}(\vr y) \rvert \leq \prod_a{\exp \left( n T_Y(a) H(\tilde{X} | \tilde{Y}=a) \right) }
=\\= \exp \left( n H(\tilde{X} | \tilde{Y}) \right)
\end{multline}
(c) is based on bounding the number of types (see \cite{Cover}, Theorem 11.1.1), and the fact that in the minimization region $I(T_Y,T_{X|Y}) \geq t$ and $D(T_X || Q) \geq 0$ therefore the result of the minimum is at least $t$.

\subsection{Discussion of Lemma 1}
\subsubsection{An alternative proof for the exponential rate}
For the proof of Theorem \ref{theorem:discrete_nonadaptive} we do not need the strict inequalities and equality in the error exponent would be sufficient, however these will be useful later for the rateless coding. An explanation for the fact that the result does not depend on $Q$ can be obtained by showing that the above probability can be bounded for each type of $\vr x$ separately. I.e. if $\vr x$ is drawn uniformly over the type $\mathcal{T}_X$ the probability of the above condition is:
\begin{multline}\label{eq:pairwise_discrete_approx}
\frac{\displaystyle \sum_{T_{XY} \in T_t} \lvert \mathcal{T}_{X|Y}(\vr y) \rvert}{\lvert \mathcal{T}_X \rvert} \doteq
 \frac{\displaystyle \sum_{T_{XY} \in T_t} \exp(n H(\tilde{X} | \tilde{Y}) )}{\exp(n H(\tilde{X}))}
=\\=
\sum_{T_{XY} \in T_t} \exp(-n I(\tilde{X} ; \tilde{Y})) \doteq \exp(-n t)
\end{multline}
where $T_t \equiv \big\{ T_{XY} \in \mathcal{P}_n(\mathcal{XY}): (T_{XY})_X = T_X, (T_{XY})_Y = T_Y, I(T_Y,T_{Y|X}) \geq t \big\} $
and since drawing $\vr x \sim Q^n$ is equivalent to first drawing the type of $\vr x$ and then drawing $\vr x$ uniformly over the type, the bound holds when $\vr x \sim Q^n$.

\subsubsection{Extension to alpha receivers}
Following we discuss an extension of the bound and relate it to Agarwal's \cite{Agarwal_RD} coding theorem using the rate distortion function. Consider a communication system similar to that of Theorem \ref{theorem:discrete_nonadaptive}, where the codebook is a constant composition code, consisting of randomly selected sequences of type $Q$, and the receiver is an $\alpha$ receiver (see \cite{CsiszarNarayan_mismatch95}), i.e. selects the received codeword by maximizing a function $\hat\alpha(\vr x, \vr y)$ depending only on the joint empirical distribution of the sequences $\vr x, \vr y$. The function $\alpha(\tilde{X}, \tilde{Y}) = \alpha(T_{XY})$ is defined as the respective function of the distribution of $\tilde{X}, \tilde{Y}$. Then, the pairwise error probability may be bounded similarly to eq. (\ref{eq:pairwise_discrete_approx}) by replacing the condition the condition $I(T_Y,T_{Y|X}) \geq t$ in the definition of $T_t$ by $\alpha(T_{XY}) \geq t$, and obtaining:
\begin{multline}\label{eq:generalized_pairwise_discrete_approx}
\Pr(\hat \alpha(\vr x, \vr y) \geq t) \leq P_\alpha
\doteq \\ \doteq \exp \bigg[ -n \bigg(\min_{ \begin{array}{c} \scriptstyle P_{\tilde{X}\tilde{Y}}: \tilde{X} \sim Q \\ \scriptstyle \tilde{Y} \sim \hat P(\vr y) \\ \scriptstyle \alpha(\tilde{X}, \tilde{Y})\geq t \end{array}} I(\tilde{X} ; \tilde{Y}) \bigg) \bigg]
\leq \\ \leq \exp \bigg[ -n \bigg(\min_{\begin{array}{c} \scriptstyle P_{\tilde{X}\tilde{Y}}: \tilde{X} \sim Q \\ \scriptstyle \alpha(\tilde{X}, \tilde{Y})\geq t \end{array}} I(\tilde{X} ; \tilde{Y}) \bigg) \bigg]
\end{multline}
Following the proof of Theorem \ref{theorem:discrete_nonadaptive}, the RHS of eq.(\ref{eq:generalized_pairwise_discrete_approx}) determines the following achievable rate:
\begin{equation}\label{eq:rate_from_pairwise_generalized_approx}
\Remp(\vr x, \vr y) \approx \min_{\begin{array}{c} \scriptstyle \tilde{X} \sim Q,\\ \scriptstyle \alpha(\tilde{X}, \tilde{Y})\geq \hat \alpha(\vr x, \vr y) \end{array}} I(\tilde{X} ; \tilde{Y}) \hspace{2ex} \stackrel{\approx}{\leq} \hat I(\vr x, \vr y)
\end{equation}
Where the approximate inequality stems from substituting the empirical distribution of $\vr x, \vr y$ as a  particular distribution of $\tilde{X}, \tilde{Y}$ meeting the minimization constraints. The above expression is similar to the one obtained in mismatch decoding with random codes. Eq.(\ref{eq:generalized_pairwise_discrete_approx}) allows a larger (but still limited) scope of empirical rate functions, but also shows that within this scope the best function is still the empirical mutual information. On the other hand, an advantage of this expression is that under some continuity conditions it can be extended from discrete to continuous vectors (as performed in \cite{Agarwal_RD}).

When substituting $\alpha$ with the distortion function $\alpha(\tilde{X}, \tilde{Y}) = - E d(\tilde{X}, \tilde{Y})$, we would obtain:
\begin{multline}\label{eq:rate_distortion_from_pairwise_generalized_approx}
\Remp(\vr x, \vr y) \approx \min_{\begin{array}{c} \scriptstyle \tilde{X} \sim Q,\\ \scriptstyle E d(\tilde{X}, \tilde{Y})\geq \hat E d(\vr x, \vr y) \end{array}} I(\tilde{X} ; \tilde{Y}) \hspace{2ex}
=\\= R_X(\hat E d(\vr x, \vr y)) = R_X(\hat D)
\end{multline}
where $R_X(D)$ is the rate distortion function of an i.i.d. source $X \sim Q$ with the distortion metric $d$. The later relation can be used to show the result that communication at the rate $R_X(D)$ is possible where $D$ is the empirical or the maximum guaranteed distortion of the channel as shown in \cite{Agarwal_RD}. On the other hand, when using the correlation function $\alpha(\tilde{X}, \tilde{Y}) = \frac{E (\tilde{X}\tilde{Y})}{E (\tilde{X}^2) E(\tilde{Y}^2)} = \rho$, we would obtain from eq.(\ref{eq:rate_from_pairwise_generalized_approx}) and Lemma \ref{lemma:gaussian_mi_bound}:  $\Remp(\vr x, \vr y) \approx - \frac{1}{2} \log(1 - \hat\rho^2)$. Note that although the later expression is the same as the one obtained in Theorem \ref{theorem:continuous_nonadaptive}, the above derivation only proves it for discrete vectors.

\subsection{Proof of Lemma 2}\label{appendix:gaussian_mi_bound}
For random variables $X$ and $Y$ where $X$ is continuous (not necessarily Gaussian) we have the following bound on the conditional differential entropy ($\tilde{Y}$ denotes a dummy variable with the same distribution as $Y$ and used for notational purposes):
\begin{multline}\label{eq:gaussian_mi_bound_eq1}
h(X|Y) = E_{\tilde{Y}} \left[ h \left(X \big\vert Y=\tilde{Y} \right) \right]
\leq \\ \stackrel{(a)}{\leq}
E \left[ \half \log \left( 2\pi e VAR(X|Y) \right) \right]
\leq \\ \stackrel{(b)}{\leq}
\half \log \left(2\pi e E \left[VAR(X|Y) \right] \right)
=\\= \half \log \left(2\pi e E \left[VAR(X - \alpha \cdot Y|Y)\right] \right)
\leq \\ \stackrel{(c)}{\leq}
\half \log \left(2\pi e E(X - \alpha \cdot Y)^2 \right) =_{\alpha:=\frac{E(XY)}{E(Y^2)}} \\
= \half \log \left(2\pi e \left(E(X^2) - \frac{E(XY)^2}{E(Y^2)} \right) \right)
=\\= \half \log \left(2\pi e E(X^2) (1 - \rho^2)\right)
=\\= \half \log \left(2\pi e E(X^2)\right) + \half \log \left(1 - \rho^2\right)
\end{multline}
where the (a) is based on Gaussian bound for entropy and (b) on concavity of the $\log$ function (see also \cite{Cover} Eq.(17.24)) (c) is based on $VAR(X) = E(X^2)-(E X)^2 \leq E(X^2)$ and is similar to the assertion that $E[VAR(X|Y)]$ which is the MMSE estimation error is not worse than the LMMSE estimation error (except our disregard for the mean).

Therefore for a Gaussian $X$:
\begin{multline}
I(X;Y) = h(X) - h(X|Y)
=\\= \half \log(2\pi e E(X^2)) - h(X|Y) \stackrel{(\ref{eq:gaussian_mi_bound_eq1})}{\geq} -\half \log(1 - \rho^2)
\end{multline}
\myendofproof

\textit{Proof of corollary \ref{corollary1_gaussian_mi_bound}}: Equality (a) holds only if $X|Y$ is Gaussian for every value of $Y$, (b) holds if $X$ has fixed variance conditioned on every $Y$, and (c) if $E(X- \alpha \cdot Y|Y)=0 \Longrightarrow E(X|Y)=\alpha \cdot Y$, therefore it results in $X|Y \sim \Normal(\alpha Y, \const)$ which implies $X,Y$ are jointly Gaussian (easy to check by calculating the pdf).

Note that if $X,Y$ are jointly Gaussian then $Y$ can be represented as a result of an additive white Gaussian noise channel (AWGN) with gain operating on $X$:
\begin{equation}
Y \sim E(Y|X) + \Normal(0,VAR(Y|X)) = \tilde{\alpha} \cdot X + \Normal(0,\sigma^2) + \const
\end{equation}

To show corollary \ref{corollary2_gaussian_mi_bound} consider $X=Y=Ber(\half)$, in which case $I(X;Y)=1$ and $\rho=1$, therefore the assertion doesn't hold.

\subsection{Proof of Lemma 4}\label{appendix:pairwise_continuous}
Write the empirical correlation as
\begin{equation}
\hat\rho \equiv \frac{\vr x^T \vr y}{\lVert \vr x \rVert \lVert \vr y \rVert } =
 \left( \frac{\vr x}{\lVert \vr x \rVert} \right) ^T \left( \frac{\vr y}{\lVert \vr y \rVert}\right)
\end{equation}
From the expression above we can infer that $\hat\rho$ does not depend on the
amplitude of $\vr x$ and $\vr y$ but only on their direction. Since $\vr x$ is
isotropically distributed, the result does not depend on the direction of $\vr
y$ (unless $\vr y = 0$ in which case it is trivially correct), therefore it is independent of $\vr y$ and we can conveniently choose $\vr y = (1,0,0,\ldots,0)$. To put the claim above more formally, for any unitary $n \times n$ matrix $\mt U$ we can write:
\begin{multline}
\hat\rho = \frac{\vr x^T \vr y}{\sqrt{(\vr x^T \vr x)(\vr y^T \vr y)}} =
\frac{\vr x^T \mt U^T \mt U \vr y}{\sqrt{(\vr x^T \mt U^T \mt U \vr x)(\vr y^T \mt U^T \mt U \vr y)}}
=\\= \left( \frac{\mt U \vr
x}{\lVert \mt U \vr x \rVert}\right)^T \left(\frac{\mt U \vr y}{\lVert \mt U
\vr y \rVert}\right)
\end{multline}
Since $\vr x$ is Gaussian, $\mt U \vr x$ has the same distribution of $\vr x$, thus the probability remains unchanged if we remove $\mt U$ from the left side and remain with $\hat\rho' = \left( \frac{\vr
x}{\lVert \vr x \rVert}\right)^T \left(\frac{\mt U \vr y}{\lVert \mt U
\vr y \rVert}\right) $. For $\vr y \neq 0$, we may choose the unitary matrix $\mt U$
whose first row is $\frac{\vr y}{\lVert \vr y \rVert}$ and the other rows
complete it to an orthonormal basis of the linear space $\mathbb{R}^n$. Then $\mt
U \vr y = (\lVert \vr y \rVert,0,0,..0)$ and therefore $\left(\frac{\mt U \vr y}{\lVert \mt U \vr y
\rVert}\right) = (1,0,0,..0)$. Thus the distribution of $\hat\rho' = (1,0,0,\ldots,0) \cdot \left( \frac{\vr
x}{\lVert \vr x \rVert}\right) = \frac{x_1}{\lVert \vr x \rVert}$ equals the distribution of $\hat\rho$.
Assuming without loss of generality that $\vr x \sim \Normal^n(0,1)$ we have:
\begin{multline}
\Pr (|\hat\rho| \geq t) = \Pr \left( \frac{x_1}{\lVert \vr x \rVert} \geq t \right)
=\\=
\Pr \left( x_1^2 \geq t^2 (\lVert \vr x_2^n \rVert^2 + x_1^2) \right)
= \\ =
\Pr \left( x_1^2 \geq \frac{t^2}{1-t^2} \lVert \vr x_2^n \rVert^2  \right)
=\\=
E \left[ \Pr \left( x_1^2 \geq \frac{t^2}{1-t^2} \lVert \vr x_2^n \rVert^2  \right) \bigg\vert \vr x_2^n \right]
=\\=
E \left[ 2 Q \left( \sqrt {\frac{t^2}{1-t^2} \lVert \vr x_2^n \rVert^2  } \right) \right] \leq
E \left[ 2 e ^{-\half \frac{t^2}{1-t^2} \lVert \vr x_2^n \rVert^2} \right]
=\\=
\int_{\mathbb{R}^{n-1}} \left( 2 e ^{-\half \frac{t^2}{1-t^2} \lVert \vr x_2^n \rVert^2} \right) \left( \frac{1}{(2 \pi)^{(n-1)/2}} e^{- \half \lVert \vr x_2^n \rVert^2} \right) d \vr x_2^n
=\\=
2 \int_{\mathbb{R}^{n-1}} \frac{1}{(2 \pi)^{(n-1)/2}} e ^{-\half \frac{1}{1-t^2} \lVert \vr x_2^n \rVert^2}
\cdot d \vr x_2^n
=\\=
2 (1-t^2)^{\frac{n-1}{2}} \int_{\mathbb{R}^{n-1}} f_{\Normal^{n-1}(0,1-t^2)}(\vr x_2^n) \cdot d \vr x_2^n
=\\=
2 (1-t^2)^{\frac{n-1}{2}} = 2 \exp \left( - (n-1) R_2(t) \right)
\end{multline}
where we used the rough upper bound of the Gaussian error function $Q(x) \equiv \Pr(\Normal(0,1)\geq x) \leq e^{-x^2/2}$, and $f_{\Normal^{n}(\mu,\sigma^2)}$ denotes the pdf of a Gaussian i.i.d. vector.
\myendofproof

\begin{figure}
  \center
  \includegraphics[width=8cm]{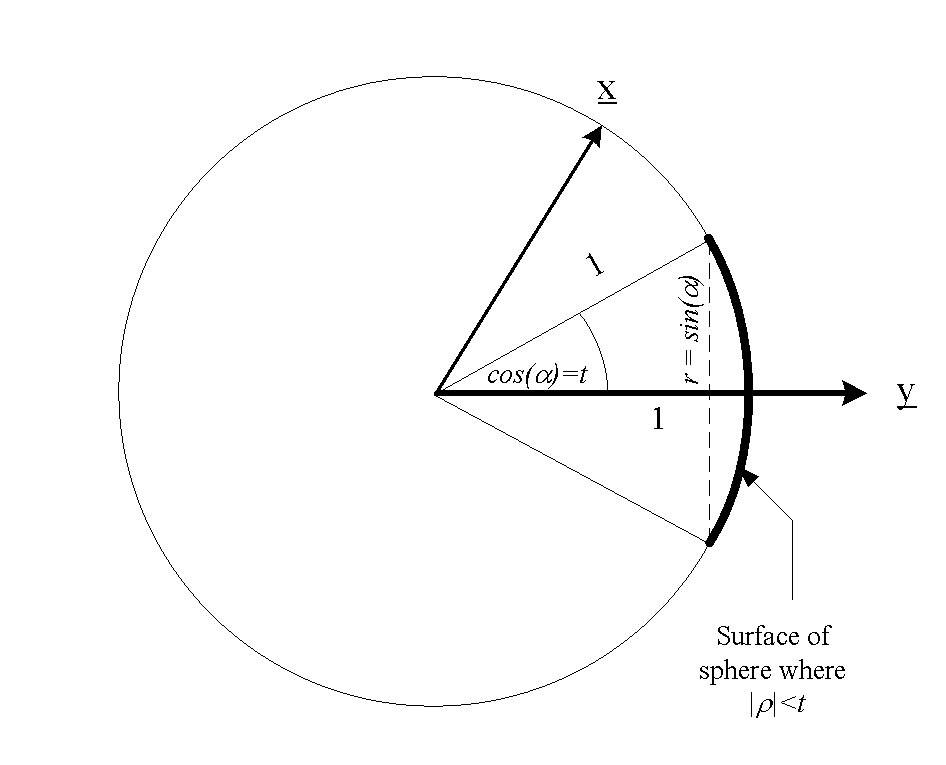}\\
  \caption{A geometric interpretation of Lemma \ref{lemma:pairwise_continuous}}\label{fig:Geometric_interpretation_of_gaussian_pairwise}
\end{figure}
\textit{Discussion:} A geometrical interpretation of Lemma \ref{lemma:pairwise_continuous} relates this probability to the
solid angle of the cone $\{\vr x: |\hat\rho| > t\}$. Since
$\vr x$ is isotropically distributed, the probability to have $|\hat\rho| > t$
equals the relative surface determined by vectors having $|\hat\rho| > t$ on the unit $n$-ball (termed the
solid angle). Since $\hat\rho$ is the cosine of the angle between $\vr x$ and $\vr y$ the points where $|\hat\rho| > t$ generate
a cone with inner angle $2\alpha$ where $\cos(\alpha)=t$ and their intersection with the unit $n$-ball is a
spherical cap (dome), shown in figure \ref{fig:Geometric_interpretation_of_gaussian_pairwise}. We can obtain a similar bound as above using geometrical considerations. Write the volume of an $n$ dimensional ball as $V_n r^n$ where $V_n$ is a fixed factor $V_n = \frac{\pi^{n/2}}{\Gamma(1+n/2)}$ \cite{Ball:Mathworld}, and accordingly the surface of an $n$ dimensional ball is (the derivative) $n V_n r^{n-1}$, then the relative surface of the spherical cap can be computed by integrating the surfaces of the $n-1$ dimensional balls with radius $\sin(\theta)$ that have a fixed angle $\theta$ with respect to $\vr y$, and can be bounded as follows:
\begin{multline}
\Pr (|\hat\rho| \geq t) = \frac{\textrm{Surface of cap}}{\textrm{Surface of ball}}
=\\=
\frac{1}{n V_n} \cdot \int_{\theta=0}^{\alpha} (n-1) V_{n-1} \sin^{n-2}(\theta) d\theta
\leq \\ \leq
 \frac{V_{n-1}}{V_n} \cdot \sin^{n-3}(\alpha) \int_{\theta=0}^{\alpha} \sin(\theta) d\theta
=\\=
\frac{V_{n-1}}{V_n} \cdot \sin^{n-3}(\alpha) (1 - \cos(\alpha))
\leq \\ \stackrel{\alpha\leq\frac{\pi}{2}}{\leq} \frac{V_{n-1}}{V_n} \cdot \sin^{n-3}(\alpha) (1 - \cos^2(\alpha))
= \\ =
O(\sqrt{n}) \cdot \sin^{n-1}(\alpha) = O(\sqrt{n}) \cdot \sqrt{1-\cos^2(\alpha)}^{n-1}
=\\=
O(\sqrt{n}) \cdot (1-t^2)^{(n-1)/2}
\end{multline}
where the asymptotic ratio $\frac{V_{n-1}}{\sqrt{n} V_n} \to 1$ is based on \cite{Gamma:Mathworld} Eq.(99). An interesting observation is that the assumption of Gaussian distribution is not necessary and this bound is true for all isotopical distributions.

\begin{table*}[t]
\caption{Parameters of adaptive rate scheme used for figure \ref{fig:rate_in_continuous_theorem}}\label{table:rate_in_continuous_theorem_params}
\center
\begin{tabular}{|p{3cm}|p{3cm}|p{4cm}|p{4cm}|}
\hline
Item            &   Referrence               &   Parameter set 1 of figure \ref{fig:rate_in_continuous_theorem} & Parameter set 2
\\ \hline Transmission scheme & section \ref{sec:rate_adaptive_scheme} &
$n=1e+008, K=1e+006, P_A=0.001, P_e=0.001$ &
$n=1e+020, K=1e+017, P_A=0.001, P_e=0.001$
\\ \hline $\RLBONE$ parameters & section \ref{sec:rate_analysis_continuous}, Eq.(\ref{eq:cont_rate_analysis9}) &
$T=2.5e+005, \Delta_{\mu} = 37.5412, \Delta = 0.0345958, \eta_1 = 0.996007, \eta_2 = 0.999962, \epsilon_1 =0.01$ &
$T=7.5e+015, \Delta_{\mu} = 77.4043, \Delta = 3.14616e-007, \eta_1 = 1, \eta_2 = 1, \epsilon_1 =0.001$
\\ \hline $\RLBTWO$ parameters & section \ref{sec:rate_analysis_continuous}, Theorem \ref{theorem:continuous_adaptive} &  $\rho_0=0.9, \epsilon=0.139438, \bar R = 1.05173$
&  $\rho_0=0.99998, \epsilon=0.0068209, \bar R = 7.29818$
\\ \hline
\end{tabular}
\end{table*}

\subsection{Proof of Lemma 6}\label{appendix:likely_convexity_of_rho}
We denote $\vr x_i, \vr y_i$ as the sub-vectors over $A_i$ (i.e.  $\vr x_i \equiv \vr x_{A_i}, \vr y_i \equiv \vr y_{A_i}$), their length by $n_i \equiv |A_i|$ and their relative length by $\lambda_i = n_i/n$.
We are interested to find a subset $J$ of $\vr x$ with bounded probability such that outside the set
$\sum_i \lambda_i \hat \rho_i^2 \geq \hat \rho^2 - \Delta$ for any $\vr y$. Consider the following inequality:
\begin{multline}\label{eq:rho_convexity_c1}
\lVert \vr x \rVert^2 \cdot \lVert \vr y \rVert^2 \cdot \hat\rho^2 = \left( \vr x^T \vr y \right)^2 =
\left( \sum_i \vr x_i^T \vr y_i \right)^2
=\\=
\left( \sum_i \hat\rho_i \lVert \vr x_i \lVert \cdot \lVert y_i \lVert \right)^2
\stackrel{(a)}{\leq}
\left( \sum_i \hat\rho_i^2 \lVert \vr x_i \lVert^2 \right)\cdot \left( \sum_i \lVert \vr y_i \lVert^2 \right)
=\\=
\left( \sum_i \lambda_i \hat\rho_i^2 + \sum_i \hat\rho_i^2 \left( \frac{\lVert \vr x_i \lVert^2}{\lVert \vr x \lVert^2} - \lambda_i  \right) \right) \cdot \lVert \vr x \lVert^2  \cdot \lVert \vr y \rVert^2
\leq \\ \stackrel{(b)}{\leq}
 \left( \sum_i \lambda_i \hat\rho_i^2  +
  \sum_i \max \left( \frac{\lVert \vr x_i \lVert^2}{\lVert \vr x \lVert^2} - \lambda_i  , 0 \right) \right) \cdot \lVert \vr x \rVert^2 \cdot \lVert \vr y \rVert^2
\end{multline}
where (a) is from Cauchy-Swartz inequality (b) is since $\hat\rho_i z_i \leq z_i$ for $z_i \geq 0$
and $\hat\rho_i z_i \leq 0$ for $z_i \leq 0$ therefore always $\hat\rho_i z_i \leq \max(z_i,0)$ (attained for
$\hat\rho_i = \Ind(z_i > 0)$). Both inequalities are tight in the sense that for each
$\vr x$ there is a sequence $\vr y$ (equivalent to choosing $\{\lVert \vr y_i \rVert^2\}$ and $\{\hat\rho_i\}$) that meets them in equality. Dividing by $\lVert \vr x \rVert^2 \cdot \lVert \vr y \rVert^2$ we have that
\begin{equation}\label{eq:rho_convexity_c2}
 \hat\rho^2 - \sum_i \lambda_i \hat\rho_i^2 \leq
  \sum_i \max \left( \frac{\lVert \vr x_i \lVert^2}{\lVert \vr x \lVert^2} - \lambda_i  , 0 \right)
\end{equation}
where the RHS depends only on $\vr x$ and should be bounded by $\Delta$. Thus the minimal set $J_{\Delta}$ is:
\begin{equation}\label{eq:rho_convexity_c2}
 J_{\Delta} \equiv \left\{ \vr x: \sum_i \max \left( \frac{\lVert \vr x_i \lVert^2}{\lVert \vr x \lVert^2} -
 \lambda_i  , 0 \right) > \Delta \right\}
\end{equation}
The set is minimal in the sense that none of its elements can be removed while meeting the conditions of the lemma.
We would like to bound the probability of $J_{\Delta}$. The result of $\sum_i \max (z_i, 0)$
is a partial sum of $z_i$, and since negative $z_i$ are not summed, it is easy to see this is the maximal
partial sum, i.e. we can write this sum alternatively as
\begin{equation}
\sum_i \max (z_i, 0) = \max_{I \in \mathcal{P}} \sum_{i \in I} z_i
\end{equation}
where $\mathcal{P} \equiv 2^{\{1,\ldots,p\}} \setminus \emptyset$  denotes all non empty sub-sets of $\{1,\ldots,p\}$, and its
size is $2^p - 1$. Therefore from the union bound we have:
\begin{multline}\label{eq:rho_convexity_c3}
 \Pr \{ J_{\Delta} \} = \Pr \left\{ \max_{I \in \mathcal{P}} \sum_{i \in I} \left( \frac{\lVert \vr x_i \lVert^2}{\lVert \vr x \lVert^2} -
 \lambda_i \right) > \Delta \right\}
\leq \\ \leq
 \sum_{I \in \mathcal{P}} \Pr \left\{ \sum_{i \in I} \left( \frac{\lVert \vr x_i \lVert^2}{\lVert \vr x \lVert^2} -
 \lambda_i \right) > \Delta \right\}
\end{multline}

To bound the above probability we first develop bound on the probability $\Pr \left( \sum_i a_i \lVert \vr x_i \rVert^2 \leq 0 \right)$ for some coefficients $a_i$:
\begin{lemma}\label{lemma:rho_convexisy_sum_bound}
Let $\vr x  \sim \Normal(0,P)^n$. For coefficients $\{a_i\}_{i=1}^p$ with $\sum_i \lambda_i a_i = \bar a > 0$ and $|a_i| \leq A$ where $|\bar a| \leq \frac{1}{8} A$, we have
\begin{equation}\label{eq:rho_convexisy_sum_bound}
\Pr \left( \sum_i a_i \lVert \vr x_i \rVert^2 \leq 0 \right) \leq e^{ - n E}
\end{equation}
where
\begin{equation}\label{eq:rho_convexisy_sum_bound_E}
E = \frac{\bar a^2}{6 A^2}
\end{equation}
\end{lemma}

Now we apply the bound to the events in Eq.(\ref{eq:rho_convexity_c3}):
\begin{eqnarray*}
& \displaystyle \sum_{i \in I} \left( \frac{\lVert \vr x_i \lVert^2}{\lVert \vr x \lVert^2} - \lambda_i \right) > \Delta & \\
& \Updownarrow & \\
& \displaystyle  \sum_{i \in I} \lVert \vr x_i \lVert^2 - \sum_{i \in I} \lambda_i \sum_{i=1}^p \lVert \vr x_i \lVert^2  > \Delta \sum_{i=1}^p \lVert \vr x_i \lVert^2 \\
& \Updownarrow & \\
& \displaystyle  \sum_{i=1}^p  \underbrace{ \left( \Delta +  \sum_{i \in I} \lambda_i - \Ind(i \in I) \right)}_{\equiv a_i}  \lVert \vr x_i \lVert^2 < 0 &
\end{eqnarray*}

We have:
\begin{multline}
\bar a = \sum_{i=1}^p \lambda_i a_i = \Delta \cdot \sum_{i=1}^p \lambda_i + \sum_{i \in I} \lambda_i \cdot \sum_{i=1}^p \lambda_i
-\\- \sum_{i=1}^p \Ind(i \in I)\lambda_i = \Delta
\end{multline}
And $|a_i| \leq 1+\Delta \equiv A$, therefore for $\Delta \leq 1/7$ we have $\bar a \leq \frac{1}{8} A$ and by Lemma \ref{lemma:rho_convexisy_sum_bound}:
\begin{equation}
\Pr \left\{ \sum_{i \in I} \left( \frac{\lVert \vr x_i \lVert^2}{\lVert \vr x \lVert^2} - \lambda_i \right) > \Delta \right\}
\leq e^{ - n E} \leq e^{ - n E_0}
\end{equation}
where
\begin{equation}
E = \frac{\bar a^2}{6 A^2} = \frac{\Delta^2}{6 (1+\Delta)^2} \geq \frac{\Delta^2}{6 (1+1/7)^2} \geq \frac{\Delta^2}{8} \equiv E_0
\end{equation}
and from Eq.(\ref{eq:rho_convexity_c3}) we have:
\begin{multline}\label{eq:rho_convexity_c4}
 \Pr \{ J_{\Delta} \} \leq
 \sum_{I \in \mathcal{P}} \Pr \left\{ \sum_{i \in I} \left( \frac{\lVert \vr x_i \lVert^2}{\lVert \vr x \lVert^2} -
 \lambda_i \right) > \Delta \right\}
\leq \\ \leq
|\mathcal{P}| \cdot e^{ - n E_0} \leq 2^p e^{ - n E_0}
\end{multline}
which proves the lemma. Note that different bounds can be obtained by applying the bound on $m$ smaller sets in $\{1,\ldots,p\}$ and requiring that the sum over each set will be bounded by $\Delta/m$ (as an example we could bound each $\max(z_i,0)$ separately by $\Delta/p$), however this bound is most suitable for our purpose since when $p << n$ the element $2^p$ becomes negligible.
\myendofproof

\textit{Proof of Lemma \ref{lemma:rho_convexisy_sum_bound}}:
We assume without loss of generality that $\vr x \sim \Normal(0,1)^n$. For Gaussian r.v. $X \sim \Normal(0,1)$ and $a < \half$ we have:
\begin{multline}
 E(e^{a x^2}) = \int_{-\infty}^{\infty} \frac{1}{\sqrt{2\pi}} e^{(a-\half)x^2} dx
=\\=
\frac{1}{\sqrt{1-2a}} \int_{-\infty}^{\infty} \frac{1}{\sqrt{2\pi (1-2a)^{-1}}} e^{-\frac{x^2}{2(1-2a)^{-1}}} dx
=\\=
\frac{1}{\sqrt{1-2a}}
\end{multline}

For coefficients $\{a_i\}_{i=1}^p$ with $\sum_i \lambda_i a_i = \bar a > 0$ and $|a_i| \leq A$, $w>0$ a positive constant of our choice, and $\vr x \sim \Normal(0,1)^n$ we have:
\begin{multline}
\ln \Pr \left( \sum_i a_i \lVert \vr x_i \rVert^2 \leq 0 \right) \leq
\ln E e^{-\half w \cdot \sum_i a_i \lVert \vr x_i \rVert^2} =\\=
\ln E e^{-\half w \cdot \sum_i a_i \sum_{j \in A_i} x_j^2}
=
\ln \prod_i \prod_{j \in A_i} E e^{-\half w \cdot a_i \cdot x_j^2}
=\\=
 \sum_i \sum_{j \in A_i} \ln \left( (1 + w \cdot a_i)^{-\half} \right)  =
-\half n \sum_i \lambda_i \ln (1+ w \cdot a_i)
= \\  \stackrel{(a)}{=}
-\half n \sum_i \lambda_i \left( (w \cdot a_i) - \half \frac{1}{(1+w\cdot t_i)^2} (w \cdot a_i)^2 \right)
\leq \\ \stackrel{(b)}{\leq}
-\half n \sum_i \lambda_i \left( (w \cdot a_i) - \half \frac{1}{(1-w \cdot A)^2} (w \cdot A)^2 \right)
= \\ =
-\half n \left( \bar a w  - \frac{A^2 w^2}{2 (1-w \cdot A)^2} \right)
\end{multline}
where (a) is based on the second order Tailor series of $\ln(1+wt)$ around $t=0$ with some $t_i \in [0,a_i] \cup [a_i,0]$ and (b) is since $|t_i| \leq |a_i| \leq A$.
For simplicity we choose a sub-optimal $w^* = \frac{\bar a}{A^2}$ (which is obtained
by assuming small $a,w$ and optimizing the bound with respect to $w$ ignoring the denominator) and obtain:
\begin{multline}
\bar a w^*  - \frac{A^2 {w^*}^2}{2 (1-w^* \cdot A)^2}
= \frac{\bar a^2}{A^2} - \frac{\bar a^2 / A^2 }{2(1-\bar a / A)^2}
=\\= \frac{\bar a^2}{A^2} \left( 1 - \frac{A^2}{2(A-\bar a)^2} \right)
\end{multline}
To simplify the bound, we make a further assumption that $|\bar a| \leq \frac{1}{8} A$ therefore:
\begin{multline}
\frac{\bar a^2}{A^2} \left( 1 - \frac{A^2}{2(A-\bar a)^2} \right)
\geq \frac{\bar a^2}{A^2} \left( 1 - \frac{A^2}{2 \cdot (7/8)^2 \cdot A^2} \right)
=\\=
\frac{\bar a^2}{A^2} \cdot \frac{17}{49}
\geq \frac{\bar a^2}{3A^2}
\end{multline}
Therefore we can write the following bound: for $|\bar a| \leq \frac{1}{8} A$ we have
\begin{equation}
\Pr \left( \sum_i a_i \lVert \vr x_i \rVert^2 \leq 0 \right) \leq e^{ - n E}
\end{equation}
where $E = \frac{\bar a^2}{6 A^2}$. Note that the bound is true for any $\vr x  \sim \Normal(0,P)^n$.
\myendofproof


\subsection{Parameters of adaptive rate scheme used for figure \ref{fig:rate_in_continuous_theorem}}
Table \ref{table:rate_in_continuous_theorem_params} lists two sets of parameters for the continuous alphabet adaptive rate scheme. The first set was used for the curves in figure \ref{fig:rate_in_continuous_theorem}, and the second set shows the convergence of $\epsilon, \bar R$, for higher values of $n,K$. Note that the values of $n,K$ are extremely high, and this is due to the looseness of the bounds used in the continuous case: specifically the exponent of Lemma \ref{lemma:likely_convexity_of_rho} which yields a relatively slow convergence of the ill-convexity probability in equation \ref{eq:convexity_prob_in_cont_rate_analysis}.



\end{document}